\documentclass[aps,twocolumn,showpacs]{revtex4}
\usepackage{graphicx} % 必需
\usepackage{caption}  % 可选，用于标题格式控制
\usepackage{subcaption}%多图并列
\usepackage{amsmath,color,ulem}
\usepackage{physics}
\usepackage{epsfig}
\usepackage{multirow}
\usepackage{float}
\usepackage{array} % 用于调整列格式
\usepackage{rotating}
\usepackage{booktabs} % 用于美化表格线
\usepackage{makecell}

\begin{document}

\title{Systematic study of baryon-baryon interactions in singly bottomed systems}
\author{Yuxuan Du$^1$}
\author{Yanyue Pan$^1$}
\author{Xinmei Zhu$^2$}
\author{Zhiyun Tan$^3$}
\author{Hongxia Huang$^1$}\email[E-mail: ]{hxhuang@njnu.edu.cn (Corresponding author)}
\author{Jialun Ping$^1$}
\affiliation{$^1$Department of Physics, Nanjing Normal University, Nanjing, Jiangsu 210097, China}
\affiliation{$^2$Department of Physics, Yangzhou University, Yangzhou 225009, People's Republic of China}
\affiliation{$^3$School of Physics and Electronic Science, Zunyi Normal College, Zunyi, Guizhou 563006, China}

	\begin{abstract}
		In this work, we systematically investigate the baryon-baryon interactions in singly bottomed dibaryon systems within the chiral quark model and search for possible bound states. By exploring the baryon-baryon interaction, we find that low-isospin channels tend to generate deeper attractive interactions and are prone to form bound states. We also find that decuplet-decuplet systems tend to show stronger attraction and support a larger number of bound states, whereas octet-octet systems generally exhibit weaker attraction and fewer bound solutions. The binding behavior of octet-decuplet systems generall lies between these two cases.
   Several bound states are obtained, which are $\Delta\Sigma_b^*$ with $IJ=\frac{1}{2}0$, $\frac{1}{2}1$, $\frac{1}{2}2$, $\frac{1}{2}3$, and $\frac{3}{2}3$, the $\Delta\Sigma_b$ with $IJ=\frac{1}{2}1$, $\frac{1}{2}2$, $\frac{3}{2}1$, and $\frac{3}{2}2$, the $N\Sigma_b^*$ with $IJ=\frac{1}{2}2$, the $\Sigma\Sigma_b$ with $IJ=00$, $01$, and $11$, the $\Sigma\Sigma_b^*$ with $IJ=01$ and $02$, the $\Sigma^*\Sigma_b$ with $IJ=01$ and $02$, and the $\Sigma^*\Sigma_b^*$ with $IJ=00$, $01$, $02$, $03$, and $13$. Further investigations of the corresponding scattering processes are still required to determine whether these bound-state candidates can be identified. 
  These states deserve further experimental investigation.
 %By exploring the baryon-baryon interaction, we find that systems composed of two decuplet baryons exhibit stronger attraction and are more favorable for forming bound states, while low-isospin channels tend to generate deeper attractive interactions and are also prone to forming bound states. 

	\end{abstract}

	\maketitle
	
	\setcounter{totalnumber}{5}
	
	\section{\label{sec:introduction}Introduction}
	
	The investigation of dibaryon (or hexaquark) states within hadron physics has a long and complicated history \cite{hc}. Dibaryon candidates composed of light quarks, such as H-dibaryon \cite{rlg,pjm,apb,mo,pes}, $d^*$ \cite{mb,pa,paw,paww,pawwm,fjd,jtg,jlp,ag,h}, $N\Omega$ \cite{jt,mok,fw,hrp,qbl,dlr,dz,ts,fe,fea,tia,ja,ac,yy} and $\Omega\Omega$ \cite{zy,zyz,lr,my,xhc}, have attracted extensive attention. The $d^*$ is an important nonstrange dibaryon resonance observed by the CELSIUS-WASA Collaboration, while the H-dibaryon provides a classic theoretical benchmark for strange six-quark systems. Studies of the $N\Omega$ and $\Omega\Omega$ systems based on quark models, lattice QCD, and correlation function analyses have also suggested possible bound state behavior.
These representative candidates have stimulated continuous interest in the search for six-quark configurations. A natural extension is therefore to explore whether similar structures can occur in systems containing heavy quarks, especially bottom quarks.

	 Experimentally, the bottom baryon sector has also made substantial progress.
	  The CDF Collaboration reported the observation of the heavy baryons $\Sigma_{b}^{\pm}$ and $\Sigma_{b}^{*\pm}$ in 2007 \cite{tae}, and subsequently observed the $\Omega_{b}^{-}$ through the decay chain $\Omega_{b}^{-} \rightarrow J/ \psi \Omega^{-}$ using 4.2 fb$^{-1}$ of data from $p \bar{p}$ collisions at $\sqrt{s}=1.96$ TeV \cite{ta}.  In recent years, several excited $\Lambda_{b}^0$ states have been discovered \cite{rae1,rae2}. These observations have encouraged experimental collaborations to explore the less known excited \(\Xi_b\) states, as $\Lambda_{b}^0$ and $\Xi_{b}$ have similar properties due to the approximate $SU(3)$ flavor symmetry. Recently, LHCb Collaboration reported the observation of $\Xi_{b}(6227)^{-}$ \cite{rae3} and $\Xi_{b}(6227)^{0}$ \cite{rae4}, and the CMS Collaboration reported the observation of the $\Xi_{b}(6100)^{-}$ \cite{ams}.
	  
	  Prompted by the aforementioned experimental studies of heavy baryons, it is natural to expand investigation to dibaryons composed of heavy baryons.
	 The main motivation for these studies originates from the reduction of the kinetic energy, which results from the substantially larger reduced mass compared with light-baryon systems. Currently, various theoretical methods have been employed to study dibaryons containing heavy quarks. These approaches include lattice QCD \cite{pja,bd,yl,pj,nm,pm,pmj}, quark model \cite{sp,jlb,dbl,jl,sm,jv,ap,jm,zcx,xz,pmh}, meson exchange model \cite{nl,nlz,rc,mz}, and so on \cite{jx,oa}.
	% In Ref. \cite{yl}, Lyu $et~al$. employed lattice QCD to investigate the $\Omega_{ccc}\Omega_{ccc}$ system in the $^{1}S_{0}$, finding it to be weakly bound with a binding energy of approximately 5.68 MeV. Using an identical approach, Mathur $et~al$. substituted charm quarks with bottom quarks and discovered a strongly bound $\Omega_{bbb}\Omega_{bbb}$ dibaryon in the same channel, exhibiting a binding energy of about 81 MeV \cite{nm}. A subsequent study based on the extended OBE model likewise confirmed the existence of such bound states, though it reported a substantially smaller binding energy of 6 MeV for the $\Omega_{bbb}\Omega_{bbb}$ \cite{mz}. 
	In Ref. \cite{sp}, Pepin $et~al$. investigated the stability of hexaquark systems with the quark content $uuddsb$ in the context of a chiral constituent quark model. The authors found that the heavy compact hexaquark $uuddsb$ was highly unstable, contrary to the findings of Refs. \cite{dbl} and \cite{jl}. In the latter reference, the system with $I=0,~J=2$ was bound by 13.8 MeV for the most favorable choice of the model parameters within the chromomagnetic model. Leandri $et~al$. also studied the content of $uudssb$ dibaryon with isospin $I=\frac{1}{2}$ and total angular momentum $J=2$, reporting a binding energy of 43.9 MeV for this configuration \cite{jl}. On the other hand, Park $et~al$. investigated the stability of dibaryons with $I=\frac{1}{2}$, containing two strange quarks and one heavy flavor \cite{ap}. They calculated the masses of systems for all configurations with spin $S=0,~1,~2$, and concluded that there are no stable and compact $uudssb$ dibaryon states.
	 
	 Although the above studies have greatly improved our understanding of heavy baryons and heavy-flavor dibaryon candidates, no dibaryon state containing heavy quarks has been experimentally established so far. 
    Theoretically, a systematic search for dibaryon states containing heavy quarks can be carried out. From the analysis of baryon-baryon systems, some regularities in the formation of bound states can be extracted, which could serve as a guide for experimental searches for dibaryon states. On the other hand, dibaryon states can also be investigated from the perspective of baryon-baryon interactions. Through analyzing these interactions, one can reveal which dibaryon states are prone to forming bound states. Therefore, in this work, we systematically search for dibaryon states containing a bottom quark and study the interactions between hadrons.
	 %Several bottomed dibaryon are identified, which are shown in Tables \ref{bound energy1} and \ref{bound energy2}.
	 %To illustrate the underlying mechanism for the emergence of these states, we have conducted a comprehensive analysis of baryon-baryon interactions among these particals.
	 
     The paper is organized as follows. After the introduction, the formalism of the model Hamiltonian is presented in Sec.~\ref{Quark model and resonating group method}. In Sec.~\ref{11}, we present our numerical results and analyze the baryon-baryon interactions in different isospin-spin channels, with the aim of identifying attractive channels that may provide candidates for future resonance searches. The last Sec.~\ref{22} is a brief summary.

\section{\label{Quark model and resonating group method}THE CHIRAL QUARK MODEL}
   Theoretical models are necessary for exploring their possible existence and internal dynamics. The quark models extensively used by our group have provided a successful framework for studying light and strange dibaryon candidates. It is therefore natural to extend this framework to dibaryon systems involving heavy quarks.
   In this work, we extend the chiral quark model (ChQM) \cite{xh,jvi} to investigate the corresponding baryon-baryon interactions involving a bottom quark. We also use the resonating group method (RGM) \cite{jar} and generating coordinate method (GCM) \cite{dlh} to perform dynamical calculations of the six‑quark system, and the details of the computational approach are given in Ref.~\cite{yxd}.
   %Phenomenological models remain the dominant instruments for understanding the nature of multiquark candidates observed in experiments.
%In order to anlayze baryon-baryon interactions, we use the chiral quark model (ChQM) \cite{xh,jvi}, resonating group method (RGM) \cite{jar}, and generating coordinate method (GCM) \cite{dlh}.
  In this section, we will provide a brief introduction to the model.

	The ChQM has been successfully employed to describe hadron spectra and hadron-hadron interactions. The construction of this model is based on the dynamical breaking of chiral symmetry \cite{ito}.
	In this framework, the Hamiltonian is composed of four parts: the mass term ($m_i$), the kinetic term ($\frac{p_i^2}{2m_i}$), the kinetic term of the center of mass ($T_c$), and the potential term ($V_{ij}$). More details of this model can be found in Refs. \cite{xh,jvi}.
	  Here, we present only the Hamiltonian and the parameters used.
\begin{equation}
	H=\sum_{i=1}^6\left(m_i+\frac{p_i^2}{2m_i}\right)-T_c+\sum_{i<j}V_{ij},
	\label{eq:Hamiltonian}
\end{equation}
	\begin{equation}
		V_{ij} = V^{\text{CON}}(r_{ij})+V^{\text{OGE}}(r_{ij}) + V^{\sigma}(r_{ij})
		+V^{\text{OBE}}(r_{ij}), \\
	\end{equation}
		where $m_i$ denotes the mass of $i$-th quark, the kinetic energy term in the Hamiltonian is $\frac{p_i^2}{2m_i}$, the term for the kinetic energy of the center of mass is $T_c$. The potential interaction $V_{ij}$ includes the confinement potential $V^{\text{CON}}(r_{ij})$, the one-gluon exchange potential $V^{\text{OGE}}(r_{ij})$, the $\sigma$ meson exchange potential $V^{\sigma}(r_{ij})$, and the one-boson exchange potential $V^{\text{OBE}}(r_{ij})$.~Here, we will present the expressions for each potential. The confinement potential $V^{\text{CON}}(r_{ij})$ is written as:
	\begin{equation}
		V^{\text{CON}}(r_{ij})= -a_c \boldsymbol{\lambda_i} \cdot \boldsymbol{\lambda_j} \left[ r_{ij}^2 + V_0 \right]. \label{eq:VCON} 
	\end{equation}
	Additionally, based on the asymptotic freedom property of QCD, the model includes the one-gluon exchange potential to describe the short-range interaction:
	\begin{align}
		V^{\text{OGE}}(r_{ij}) &= \frac{1}{4} \alpha_{s_{q_{i}q_{j}}} \boldsymbol{\lambda_i} \cdot \boldsymbol{\lambda_j} \Bigg[ \frac{1}{r_{ij}}  	- \frac{\pi}{2} \delta(\boldsymbol{r_{ij}})  \nonumber \\
		         & \left( \frac{1}{m_i^2} + \frac{1}{m_j^2} + \frac{4\boldsymbol{\sigma}_i \cdot \boldsymbol{\sigma}_j}{3m_i m_j} \right) - \frac{3}{4m_i m_j r_{ij}^3} S_{ij} \Bigg], \\ \label{eq:VOGE}
		         S_{ij} &= \frac{(\boldsymbol{\sigma}_i \cdot \boldsymbol{r}_{ij})(\boldsymbol{\sigma}_j \cdot \boldsymbol{r}_{ij})}{r_{ij}^2} - \frac{1}{3} \boldsymbol{\sigma}_i \cdot \boldsymbol{\sigma}_j ,
	\end{align}
   where $S_{ij}$ is the quark tensor operator. Since only S-wave states are considered in this work, the tensor force operator does not contribute. In the chiral quark model, we introduce the scalar $\sigma$ meson exchange potential (only between $u$ and $d$ quarks) to provide intermediate-range attraction. Its specific expression is
	\begin{align}
		V^{\sigma}(r_{ij}) &= -\frac{g^{2}_{\text{ch}}}{4\pi}\frac{\Lambda_{\sigma}^{2}m_{\sigma}}{\Lambda_{\sigma}^{2}-m_{\sigma}^{2}}
		\left[Y(m_\sigma r_{ij}) - \frac{\Lambda_{\sigma}}{m_{\sigma}}Y(\Lambda_{\sigma} r_{ij})\right].  \label{eq:Vsigma}
		\end{align}\\
		
  % The impact of heavy meson exchange has been systematically considered in numerous investigations of charmed mutiquark states \cite{ly, spf, fls}. %For example, in our previous work of the doubly heavy dibaryon systems with strangeness $S =0$ \cite{zcx}, we generalized the model from $SU(3)$ to $SU(4)$ and introduced $D$, $D_s$, and $\eta_{c}$ meson exchange interactions. The reason why heavy meson exchange can change the binding energy in doubly charmed dibaryon systems is that more heavy‑light pairs exist in the six‑quark system than in two separate baryons.
  Within the $SU(3)$ framework, the one-boson-exchange potential involves $\pi$, $k$, and $\eta$ exchanges. Since our system contains a bottom quark, we also incorporates additional heavy meson exchange mechanisms: $B$, $B_s$, and $\eta_{b}$ meson.
  Therefore, we extend the original $SU(3)$ to $SU(5)$ and explicitly include the corresponding interaction terms.	
	Within this expanded framework, the one-boson-exchange potential is written as follows
		\begin{align}
		V^{\text{OBE}}(r_{ij}) &=
		\nu_{\pi}(r_{ij})\sum_{a=1}^3 \lambda^{a}_i \cdot \lambda^{a}_j
		+ \nu_{K}(r_{ij})\sum_{a=4}^7 \lambda^{a}_i \cdot \lambda^{a}_j \nonumber \\
		&\quad + \nu_{\eta}(r_{ij})\left[
		\left(\lambda^{8}_i \cdot \lambda^{8}_j\right)\cos\theta_p
		- \left(\lambda^{0}_i \cdot \lambda^{0}_j\right)\sin\theta_p
		\right] \nonumber \\
		&\quad + \nu_{B}(r_{ij})\sum_{a=16}^{19} \lambda^{a}_i \cdot \lambda^{a}_j
		+ \nu_{B_s}(r_{ij})\sum_{a=20}^{21} \lambda^{a}_i \cdot \lambda^{a}_j \nonumber \\
		&\quad + \nu_{\eta_b}(r_{ij})\lambda^{24}_i \cdot \lambda^{24}_j \nonumber,  \label{eq:VOBE} \\
		\end{align}
		\begin{align}
		\nu_{\chi}(r_{ij}) &= \frac{g^{2}_{\text{ch}}}{4\pi}\frac{m_{\chi}^2}{12m_{i}m_{j}}
		\frac{\Lambda_{\chi}^2}{\Lambda_{\chi}^2-m_{\chi}^2}m_\chi \Bigg\{
		\nonumber \\
		&\left[Y(m_{\chi} r_{ij}) - \frac{\Lambda_{\chi}^3}{m_{\chi}^3}Y(\Lambda_{\chi} r_{ij})\right] \boldsymbol{\sigma}_i \cdot \boldsymbol{\sigma}_j \nonumber \\
		&\quad + \left[H(m_{\chi} r_{ij}) - \frac{\Lambda_{\chi}^3}{m_\chi^3} H(\Lambda_{\chi} r_{ij})\right] S_{ij} \Bigg\}. \label{eq:nux} 	
	\end{align}
	The symbol $\lambda^{a}$ in Eq (\ref{eq:VOBE}) denotes Gell-Mann matrices.
The functions $Y (x)$ and $H(x)$ presented in Eq (\ref{eq:nux}) are standard Yukawa functions \cite{ava}.
The following procedure is used to fix the model parameters. Mass parameters such as  $m_\pi$, $m_K$, $m_\eta$, $m_B$, $m_{B_{s}}$, and $m_{\eta_{b}}$ are taken from their experimental values. The mass of $u/d$-quark ($m_{u,d}$), cutoff parameters (e.g. $\Lambda_\pi$, $\Lambda_K$, $\Lambda_\eta$) and the mixing angle $\theta_p$ are empirically fitted~\cite{jvi}. The chiral coupling constant $g_{ch}$ can be derived from the $\pi NN$ coupling constant through
\begin{align}
\frac{g^{2}_{ch}}{4\pi}= \left(\frac{3}{5}\right)^{2}\frac{g^{2}_{\pi NN}}{4\pi}\frac{m^{2}_{u,d}}{m^{2}_{N}}.
       \end{align}
The remaining adjustable parameters are obtained by fitting the ground-state light baryons and singly heavy baryons. In quark model calculations, the spectra of baryons and mesons are described  by equivalent coupling constants. According to the expression of $V^{\text{OGE}}$, the chromomagnetic term depends on $\frac{\alpha_{s_{q_{i}q_{j}}}}{m_i m_j}$, where $\alpha_{s_{q_{i}q_{j}}}$ is associated with the quark flavor and determined by the mass difference between two baryons with spins $S=\frac{1}{2},~S=\frac{3}{2}$, respectively. Since heavy quarks possess a relatively large mass, the spin splitting between the two spin states $S=\frac{1}{2},~S=\frac{3}{2}$ is reduced.  Consequently, it is necessary to increase the value of $\alpha_{s_{q_{i}q_{j}}}$ to counteract the effect caused by the quark mass and to achieve a mass difference consistent with the experimental data. 
The model parameters and the masses of the fitted baryons are shown in Tables~\ref{Model parameters} and \ref{masses}, respectively.
%However, the bottom quark is extermely heavy, and its effect is excepted to small. We calculated the baryon and dibaryon masses with and without the considering of the additional heavy meson exchange terms, and found that impact of additional meson exchange potential is on the order of $10^{-2}$ for both baryons and dibarons.
%The energies listed in Tables \ref{tab1}-\ref{tab2} are expressed in the natural unit system ($\hbar$ = $c$ = 1).
	
\begin{table}[H]
		\centering
		\caption{Model parameters: $m_{\pi}=0.70~fm^{-1}$, $m_{K}=2.51~fm^{-1}$, $m_{\eta}=2.77~fm^{-1}$, $m_{\sigma}=3.42~fm^{-1}$, $m_{B}=26.75~fm^{-1}$, $m_{B_s}=27.20~fm^{-1}$, $m_{\eta_b}=47.63~fm^{-1}$, $\Lambda_{\pi}=\Lambda_{\sigma}=4.20~fm^{-1}$, $\Lambda_{K}=\Lambda_{\eta}=5.20~fm^{-1}$, $\Lambda_{B}=\Lambda_{B_s}=\Lambda_{\eta_b}=1.20~fm^{-1}$.}
		\label{Model parameters}
		% 第一张表格（N 到 Sigma_c）
		\begin{tabular}{c*{9}{c}}  % 去掉 | 符号
			\hline
			\hline
			& $b(fm)$ & $m_{u,d}($MeV$)$ & $m_{s}($MeV$)$ & $m_{c}($MeV$)$ & $m_{b}($MeV$)$  \\
			
			 & 0.49088 & 313 & 590 & 1700 & 5244 \\
			\hline
			 & $a_{c}($MeV$ \cdot fm^{-2})$  & $V_0(fm^{2})$ & $\alpha_{s_{qq}}$ & $\alpha_{s_{qs}}$ & $\alpha_{s_{ss}}$  \\
			  & 50.330  & -1.2779 & 0.50594 & 0.82961 & 0.74093 \\
			\hline
			  & $\alpha_{s_{qb}}$& $\alpha_{s_{sb}}$ \\
			 & 0.72087&0.99010\\
          \hline			
		\hline	
		\end{tabular}
	\end{table}
	
	\begin{table}[H]
		\centering
		\caption{The calculated masses (in MeV) of the baryons in
ChQM. Experimental values are taken from the Particle Data
Group \cite{jbw}.}
		\label{masses}
		% 第一张表格（N 到 Sigma_c）
		\begin{tabular}{c*{9}{c}}  % 去掉 | 符号
			\hline
			\hline
			& $N$ & $\Delta$ & $\Lambda$ & $\Sigma$ & $\Sigma^*$ & $\Omega$ & $\Xi$ & $\Xi^*$ & $\Lambda_b$  \\
			\hline
			\text{ChQM} & 944 & 1262 & 1067 & 1148 & 1345 & 1564 & 1249 & 1446 & 5646 \\
			\text{Exp.} & 939 & 1233 & 1116 & 1189 & 1315 & 1672 & 1385 & 1530 & 5620 \\
			\hline
			
			\hline
			& $\Sigma_b$  & $\Sigma^*_b$ & $\Xi_b$ & $\Xi'_b$ & $\Xi^*_b$  & $\Omega_b$ \\
			\hline
			\text{ChQM} & 5848 & 5863 & 5769 & 5892 & 5905 & 5953 \\
			\text{Exp.} & 5811 & 5830 & 5792 & 5935 & 5945 & 6046 \\
			\hline
			\hline
		\end{tabular}
	\end{table}

	\section{Results and Discussions}
	\label{11}
	
	%In this work, we mainly calculate the binding energies of different $S$-wave states to search for possible bound states.
	In this work, we systematically investigate singly bottom dibaryon systems using the RGM and GCM. We mainly calculate the binding energies of different $S$-wave states to search for possible bound states. The study covers all quantum numbers with isospin $I=0,~\frac{1}{2},~1,~\frac{3}{2},~2,~\frac{5}{2}$ and the total angular momentum $J=0,~1,~2,~3$. These channels are listed in Tables \ref{tab:channels_0p}-\ref{tab:channels_3p}. In order to identify the mechanism of bound states, we subsequently explore the origin of interactions between baryons and analyzed their corresponding effective potential. 
	 %To explore the origin of these bound states, we study the interactions between two baryons and use these interactions to determine whether bound states can be formed.
\begin{table}[H]
\centering
\caption{The singly bottom dibaryons with $J^P=0^+$, classified by $I$ and strangeness $S$.}
\label{tab:channels_0p}
\footnotesize
\setlength{\tabcolsep}{3pt}
\renewcommand{\arraystretch}{1.4}
\begin{tabular}{@{}c c p{0.88\linewidth}@{}}
\Xhline{1.0pt}
$I$ & $S$ & Channels \\
\Xhline{1.0pt}

\multirow{3}{*}{$0$}
& $-1$ & $\Lambda\Lambda_b,~ N\Xi_b,~N\Xi'_b,~\Sigma\Sigma_b,~ \Sigma\Sigma_b^*$ \\
& $-3$ & $\Lambda\Omega_b,~ \Xi\Xi_b,~ \Xi\Xi'_b,~ \Xi\Xi_b^*$ \\
& $-5$ & $\Omega\Omega_b^*$ \\
\cline{1-3}

\multirow{3}{*}{$\frac{1}{2}$}
& $0$ & $N\Lambda_b,~N\Sigma_b,~ \Delta\Sigma_b^*$ \\
& $-2$ & $N\Omega_b,~\Lambda\Xi_b,~\Xi\Lambda_b,~\Sigma\Xi_b,~ \Lambda\Xi'_b,~ \Sigma\Xi'_b,~ \Xi\Sigma_b,~\Sigma^*\Xi_b^*,~ \Xi^*\Sigma_b^*$ \\
& $-4$ & $\Xi\Omega_b,~\Xi^*\Omega_b^*,~ \Omega\Xi_b^*$ \\
\cline{1-3}

\multirow{2}{*}{$1$}
& $-1$ & $N\Xi_b,~\Sigma\Lambda_b,~N\Xi'_b,~\Lambda\Sigma_b,~\Sigma\Sigma_b,~\Delta\Xi_b^*,~\Sigma^*\Sigma_b^*$ \\
& $-3$ & $\Xi\Xi_b,~\Sigma\Omega_b,~\Xi\Xi'_b,~\Sigma^*\Omega_b^*,~\Xi^*\Xi_b^*,~\Omega\Sigma_b^*$ \\
\cline{1-3}

\multirow{2}{*}{$\frac{3}{2}$}
& $0$ & $N\Sigma_b,~ \Delta\Sigma_b^*$ \\
& $-2$ & $\Sigma\Xi_b,~ \Sigma\Xi'_b,~ \Xi\Sigma_b,~\Delta\Omega_b^*,~\Sigma^*\Xi_b^*,~ \Xi^*\Sigma_b^*$ \\
\cline{1-3}

$2$
& $-1$ & $\Sigma\Sigma_b,~\Delta\Xi_b^*,~\Sigma^*\Sigma_b^*$ \\
\cline{1-3}

$\frac{5}{2}$
& $0$ & $\Delta\Sigma_b^*$ \\
\Xhline{1.0pt}
\end{tabular}
\end{table}

\begin{table}[H]
\centering
\caption{The singly bottom dibaryons with $J^P=1^+$, classified by $I$ and strangeness $S$.}
\label{tab:channels_1p}
\footnotesize
\setlength{\tabcolsep}{3pt}
\renewcommand{\arraystretch}{1.4}
\begin{tabular}{@{}c c p{0.88\linewidth}@{}}
\Xhline{1.0pt}
$I$ & $S$ & Channels \\
\Xhline{1.0pt}

\multirow{3}{*}{$0$}
& $-1$ & $\Lambda\Lambda_b,~N\Xi_b,~ N\Xi'_b,~ N\Xi_b^*,~ \Sigma\Sigma_b,~ \Sigma\Sigma_b^*,~\Sigma^*\Sigma_b,~\Sigma^*\Sigma_b^*$ \\
& $-3$ & $\Lambda\Omega_b,~ \Lambda\Omega_b^*,~ \Omega\Lambda_b,~ \Xi\Xi_b,~ \Xi\Xi'_b,~ \Xi\Xi_b^*,~ \Xi^*\Xi_b,~ \Xi^*\Xi'_b,~ \Xi^*\Xi_b^*$ \\
& $-5$ & $\Omega\Omega_b,~ \Omega\Omega_b^*$ \\
\cline{1-3}

\multirow{4}{*}{$\frac{1}{2}$}
& $0$ & $N\Lambda_b,~N\Sigma_b,~ N\Sigma_b^*,~ \Delta\Sigma_b,~ \Delta\Sigma_b^*$ \\
& \multirow{2}{*}{$-2$} & $\Lambda\Xi_b,~\Xi\Lambda_b,~N\Omega_b,~N\Omega_b^*,~\Sigma\Xi_b,~ \Lambda\Xi'_b,~ \Lambda\Xi^*_b,~ \Sigma\Xi'_b,~ \Xi\Sigma_b,$ \\
& & $\Xi^*\Lambda_b,~ \Sigma\Xi_b^*,~ \Xi\Sigma^*_b,~ \Sigma^*\Xi_b,~\Sigma^*\Xi'_b,~ \Sigma^*\Xi_b^*,~ \Xi^*\Sigma_b,~ \Xi^*\Sigma_b^*$ \\
& $-4$ & $\Xi\Omega_b,~\Xi^*\Omega_b,~\Xi\Omega_b^*,~ \Omega\Xi_b,~ \Omega\Xi_b',~ \Omega\Xi_b^*,~\Xi^*\Omega_b^*$ \\
\cline{1-3}

\multirow{4}{*}{$1$}
& \multirow{2}{*}{$-1$} & $N\Xi_b,~\Sigma\Lambda_b,~ N\Xi'_b,~ N\Xi_b^*,~ \Lambda\Sigma_b,~ \Lambda\Sigma_b^*,~ \Sigma\Sigma_b,~ \Sigma^*\Lambda_b,~ \Sigma\Sigma_b^*,$ \\
&  & $\Delta\Xi_b,~\Delta\Xi_b',~ \Delta\Xi_b^*,~\Sigma^*\Sigma_b,~\Sigma^*\Sigma_b^*$ \\
& \multirow{2}{*}{$-3$} & $\Xi\Xi_b,~\Sigma\Omega_b,~\Sigma\Omega_b^*,~ \Xi\Xi'_b,~ \Xi\Xi_b^*,~ \Xi^*\Xi_b,~\Sigma^*\Omega_b,~\Sigma^*\Omega_b^*,~ \Xi^*\Xi_b',$\\
& &$\Xi^*\Xi_b^*,~\Omega\Sigma_b,~\Omega\Sigma_b^*$ \\
\cline{1-3}

\multirow{3}{*}{$\frac{3}{2}$}
& $0$ & $N\Sigma_b,~ N\Sigma_b^*,~ \Delta\Lambda_b,~ \Delta\Sigma_b,~ \Delta\Sigma_b^*$ \\
& \multirow{2}{*}{$-2$} & $\Sigma\Xi_b,~ \Sigma\Xi'_b,~ \Xi\Sigma_b,~ \Sigma\Xi_b^*,~ \Xi\Sigma^*_b,~ \Sigma^*\Xi_b,~\Delta\Omega_b,~\Delta\Omega_b^*,~\Sigma^*\Xi'_b,$ \\
& & $\Sigma^*\Xi_b^*,~\Xi^*\Sigma_b,~ \Xi^*\Sigma_b^*$ \\
\cline{1-3}

$2$
& $-1$ & $\Sigma\Sigma_b,~\Sigma\Sigma_b^*,~\Delta\Xi_b,~\Delta\Xi_b',~\Delta\Xi_b^*,~\Sigma^*\Sigma_b,~\Sigma^*\Sigma_b^*$ \\
\cline{1-3}

$\frac{5}{2}$
& $0$ & $\Delta\Sigma_b,~\Delta\Sigma_b^*$ \\
\Xhline{1.0pt}
\end{tabular}
\end{table}

\begin{table}[H]
\centering
\caption{The singly bottom dibaryons with $J^P=2^+$, classified by $I$ and strangeness $S$.}
\label{tab:channels_2p}
\footnotesize
\setlength{\tabcolsep}{3pt}
\renewcommand{\arraystretch}{1.4}
\begin{tabular}{@{}c c p{0.88\linewidth}@{}}
\Xhline{1.0pt}
$I$ & $S$ & Channels \\
\Xhline{1.0pt}

\multirow{3}{*}{$0$}
& $-1$ & $N\Xi_b^*,~ \Sigma\Sigma_b^*,~ \Sigma^*\Sigma_b,~ \Sigma^*\Sigma_b^*$ \\
& $-3$ & $\Lambda\Omega_b^*,~ \Omega\Lambda_b,~ \Xi\Xi_b^*,~ \Xi^*\Xi_b,~ \Xi^*\Xi'_b,~ \Xi^*\Xi_b^*$ \\
& $-5$ & $\Omega\Omega_b,~ \Omega\Omega_b^*$ \\
\cline{1-3}

\multirow{3}{*}{$\frac{1}{2}$}
& $0$ & $N\Sigma_b^*,~ \Delta\Sigma_b,~ \Delta\Sigma_b^*$ \\
& $-2$ & $N\Omega_b^*,~ \Lambda\Xi^*_b,~\Sigma\Xi_b^*,~ \Xi\Sigma^*_b,~ \Xi^*\Lambda_b,~ \Sigma^*\Xi_b,~ \Sigma^*\Xi'_b,~ \Sigma^*\Xi_b^*,~ \Xi^*\Sigma_b,$ \\
& $-4$ & $\Xi^*\Sigma_b^*,~\Xi^*\Omega_b,~\Xi\Omega_b^*,~ \Omega\Xi_b,~ \Omega\Xi_b',~ \Omega\Xi_b^*,~\Xi^*\Omega_b^*$ \\
\cline{1-3}

\multirow{2}{*}{$1$}
& $-1$ & $N\Xi_b^*,~ \Lambda\Sigma_b^*,~ \Sigma^*\Lambda_b,~ \Sigma\Sigma_b^*,~ \Delta\Xi_b,~ \Delta\Xi_b',~ \Delta\Xi_b^*,~ \Sigma^*\Sigma_b,~ \Sigma^*\Sigma_b^*$ \\
& $-3$ & $\Sigma\Omega_b^*,~ \Xi\Xi_b^*,~ \Xi^*\Xi_b,~\Sigma^*\Omega_b,~\Sigma^*\Omega_b^*,~ \Xi^*\Xi'_b,~ \Xi^*\Xi^*_b,~\Omega\Sigma_b,~ \Omega\Sigma_b^*$ \\
\cline{1-3}

\multirow{2}{*}{$\frac{3}{2}$}
& $0$ & $N\Sigma_b^*,~ \Delta\Lambda_b,~ \Delta\Sigma_b,~ \Delta\Sigma_b^*$ \\
& $-2$ & $\Sigma\Xi_b^*,~ \Xi\Sigma^*_b,~ \Sigma^*\Xi_b,~\Delta\Omega_b,~\Delta\Omega_b^*,~\Sigma^*\Xi'_b,~ \Sigma^*\Xi_b^*,~\Xi^*\Sigma_b,~ \Xi^*\Sigma_b^*$ \\
\cline{1-3}

$2$
& $-1$ & $\Sigma\Sigma_b^*,~\Delta\Xi_b,~\Delta\Xi_b',~\Delta\Xi_b^*,~\Sigma^*\Sigma_b,~\Sigma^*\Sigma_b^*$ \\
\cline{1-3}

$\frac{5}{2}$
& $0$ & $\Delta\Sigma_b,~\Delta\Sigma_b^*$ \\
\Xhline{1.0pt}
\end{tabular}
\end{table}

\begin{table}[H]
\centering
\caption{The singly bottom dibaryons with $J^P=3^+$, classified by $I$ and strangeness $S$.}
\label{tab:channels_3p}
\footnotesize
\setlength{\tabcolsep}{3pt}
\renewcommand{\arraystretch}{1.4}
\begin{tabular}{@{}c c p{0.88\linewidth}@{}}
\Xhline{1.0pt}
$I$ & $S$ & Channels \\
\Xhline{1.0pt}

\multirow{3}{*}{$0$}
& $-1$ & $\Sigma^*\Sigma_b^*$ \\
& $-3$ & $\Xi^*\Xi_b^*$ \\
& $-5$ & $\Omega\Omega_b^*$ \\
\cline{1-3}

\multirow{3}{*}{$\frac{1}{2}$}
& $0$ & $\Delta\Sigma_b^*$ \\
& $-2$ & $\Sigma^*\Xi_b^*,~ \Xi^*\Sigma_b^*$ \\
& $-4$ & $\Xi^*\Omega_b^*,~ \Omega\Xi_b^*$ \\
\cline{1-3}

\multirow{2}{*}{$1$}
& $-1$ & $\Delta\Xi_b^*,~ \Sigma^*\Sigma_b^*$ \\
& $-3$ & $\Sigma^*\Omega_b^*,~\Xi^*\Xi_b^*,~\Omega\Sigma^*_b$ \\
\cline{1-3}

\multirow{2}{*}{$\frac{3}{2}$}
& $0$ & $\Delta\Sigma_b^*$ \\
& $-2$ & $\Delta\Omega_b^*,~\Sigma^*\Xi_b^*,~ \Xi^*\Sigma_b^*$ \\
\cline{1-3}

$2$
& $-1$ & $\Delta\Xi_b^*,~\Sigma^*\Sigma_b^*$ \\
\cline{1-3}

$\frac{5}{2}$
& $0$ & $\Delta\Sigma_b^*$ \\
\Xhline{1.0pt}
\end{tabular}
\end{table}

\subsection{Bound state calculation}
\label{bound}

 We numerically calculate the binding energies $B_{sc}$   for all possible states, where $B_{sc}=E_{sc}-E_{{th}}$.  Here $E_{sc}$ represents the channel energy, and $E_{th}$ denotes the corresponding theoretical threshold. Our calculations show that most of these states are unbound with energies above the corresponding thresholds. To save space, we do not present the states that are unbound for the all quantum numbers. For those that are bound only in certain quantum numbers, we list the numerical results in Tables \ref{bound energy1}-\ref{bound energy2}. %In these Tables, "ub" stands for an unbound state, while a dash "–" marks a channel forbidden by the relevant quantum numbers.
 We now analyze the bound states summarized in these Tables.

\begin{table*}
\centering
\small
\renewcommand{\arraystretch}{1.25}
\setlength{\tabcolsep}{4pt}
\caption{Binding energies of $\Delta\Sigma^*_b$, $\Delta\Sigma_b$, and $N\Sigma^*_b$ for different $J$ and $I$ (units: MeV). Here, "ub" stands for an unbound state, while a dash "–" marks a channel forbidden by the relevant quantum numbers.}
\label{bound energy1}
\begin{tabular}{l *{12}{c}}
\Xhline{1.0pt}
& \multicolumn{3}{c}{$J=0$}
& \multicolumn{3}{c}{$J=1$}
& \multicolumn{3}{c}{$J=2$}
& \multicolumn{3}{c}{$J=3$} \\
\cline{2-13}
& $I=\frac{1}{2}$ & $I=\frac{3}{2}$ & $I=\frac{5}{2}$~~~~
& $I=\frac{1}{2}$ & $I=\frac{3}{2}$ & $I=\frac{5}{2}$~~~~
& $I=\frac{1}{2}$ & $I=\frac{3}{2}$ & $I=\frac{5}{2}$~~~~
& $I=\frac{1}{2}$ & $I=\frac{3}{2}$ & $I=\frac{5}{2}$~~~~\\
\Xhline{1.0pt}
$\Delta\Sigma^*_b$
& -32.93 & ub & ub
& -38.81 & ub & ub
& -36.96 & ub & ub
& -35.93 & -12.25 & ub \\
$\Delta\Sigma_b$
& - & - & -
& -38.43 & -0.31 & ub
& -28.68 & -1.55 & ub
& - & - & - \\
$N\Sigma^*_b$
& - & - & -
& ub & ub & -
& -1.59 & ub & -
& - & ... & - \\
\Xhline{1.0pt}
\end{tabular}
\end{table*}

\begin{table*}
\centering
\small
\renewcommand{\arraystretch}{1.25}
\setlength{\tabcolsep}{4pt}
\caption{Binding energies of $\Sigma\Sigma_b$, $\Sigma\Sigma^*_b$, $\Sigma^*\Sigma_b$, and $\Sigma^*\Sigma^*_b$ for different $J$ and $I$ (units: MeV). Here, "ub" stands for an unbound state, while a dash "–" marks a channel forbidden by the relevant quantum numbers.}
\label{bound energy2}
\begin{tabular}{l *{12}{c}}
\Xhline{1.0pt}
& \multicolumn{3}{c}{$J=0$}
& \multicolumn{3}{c}{$J=1$}
& \multicolumn{3}{c}{$J=2$}
& \multicolumn{3}{c}{$J=3$} \\
\cline{2-13}
& $I=0$ & $I=1$ & $I=2$~~~~
& $I=0$ & $I=1$ & $I=2$~~~~
& $I=0$ & $I=1$ & $I=2$~~~~
& $I=0$ & $I=1$ & $I=2$~~~~ \\
\Xhline{1.0pt}
$\Sigma\Sigma_b$
& -43.13 & ub & ub
& -18.22 & -0.03 & ub
& - & - & -
& - & - & - \\
$\Sigma\Sigma^*_b$
& - & - & -
& -30.58 & ub & ub
& -21.07 & ub & ub
& - & - & - \\
$\Sigma^*\Sigma_b$
& - & - & -
& -21.37 & ub & ub
& -18.20 & ub & ub
& - & - & - \\
$\Sigma^*\Sigma^*_b$
& -33.68 & ub & ub
& -24.16 & ub & ub
& -19.62 & ub & ub
& -24.21 & -2.27 & ub \\
\Xhline{1.0pt}
\end{tabular}
\end{table*}

%Using the quark model and computational methods described in the previous section, we numerically calculate the binding energies $B_{sc}$ of each single channel for all possible molecular states, where $B_{sc}=E_{sc}-E_{{th}}$.  Here $E_{sc}$ represents the single‑channel energy, and 
%$E_{th}$ denotes the corresponding theoretical threshold. States that are not bound are denoted by "ub" (unbound). Channels that are not allowed under the relevant quantum numbers are denoted by a dash "-". Due to the large number of possible baryon-baryon channels, the complete list of channels considered in the present paper is provided in Appendix \ref{app}. Several singly bottomed dibaryon states are obtained and presented in Tables \ref{bound energy1} and \ref{bound energy2}.
 
     %From Tables \ref{bound energy1} and \ref{bound energy2}, one can see that the $\Delta\Sigma_b^*$, $\Delta\Sigma_b$, $\Sigma^*\Sigma_b^*$, $\Sigma^*\Sigma_b$, $\Sigma\Sigma_b^*$ and $\Sigma\Sigma_b$ systems more likely to form bound states. 
     Our analysis of fully light dibaryon systems reveals that the interaction between two decuplet baryons is sufficiently attractive to form bound states, while that between two octet baryons is weak or even repulsive. The attraction between a decuplet and an octet baryons lies in between these two cases \cite{mc}.
     In our investigation of singly bottom dibaryon systems, we observe that certain states exhibit the same regularities. For example, we find that $\Delta\Sigma_b^*$ with quantum numbers $IJ=\frac{1}{2}0,~\frac{1}{2}1,~\frac{1}{2}2$, and $\frac{1}{2}3$ support several deeply bound states, especially in the low-isospin channels, as listed in Table \ref{bound energy1}. Meanwhile, $\Sigma^*\Sigma_b^*$ with quantum numbers $IJ=00,~01,~02$, and $03$
also have bound states in the low-isospin channels, which displayed in the Table \ref{bound energy2}.~Channels formed by one decuplet baryon and one octet baryon can also exhibit attractive interactions, but the attraction is generally not as strong as that in the decuplet-decuplet channels. Typical examples are the $\Delta\Sigma_b$ and $\Sigma^*\Sigma_b$ channels. The $\Delta\Sigma_b$ channel exhibits bound states with $IJ=\frac{1}{2}1$ and $\frac{1}{2}2$. The $\Sigma^*\Sigma_b$ channel support bound states with $IJ=01$ and $02$.
   However, the $\Sigma\Sigma_b^*$ channel with $IJ=01$ provides a notable exception among the octet-decuplet channels, since it also gives relatively deeply bound solutions in the low-isospin channels. Channels composed of two octet baryons generally show weaker attraction and are less likely to form bound states. However, this tendency is also not universal. An exceptional case is the $\Sigma\Sigma_b$ system, where a deeply bound state is obtained in the $IJ=00$. 
   From these Tables, we can also find that decuplet-decuplet systems appear to support a relatively larger number of bound states, suggesting that such configurations may be more favorable for binding. In contrast, octet-octet systems tend to fewer bound states, while octet-decuplet systems generally show an intermediate behavior between these two cases.
   Therefore, the baryon multiplet structure provides useful guidance for understanding the binding behavior, but the final formation of a bound state still depends on the combined effects of isospin, total spin, and the detailed baryon-baryon interaction.

     From Tables \ref{bound energy1} and \ref{bound energy2}, we can further examine the dependence of the binding energies on different quantum numbers.
     Since only ground state $S$-wave systems are considered in present work, the orbital angular momentum is zero, and the total angular momentum $J$ is determined by the total spin. For a fixed $J$, the magnitude of the binding energy generally decreases as $I$ increases, and the corresponding states eventually become unbound. Taking the $\Delta\Sigma_b^*$ system as an example, in the $J=0$ system, the state with $I=\frac{1}{2}$ has a binding energy of $-32.93$ MeV, while the states with $I=\frac{3}{2}$ and $I=\frac{5}{2}$ are unbound. Similar behavior is found for $J=1$ and $J=2$, where only the $I=\frac{1}{2}$ system remains bound, with binding energies of $-38.81$ MeV and $-36.96$ MeV, respectively. For $J=3$, the $I=\frac{1}{2}$ state is bound by $-35.93$ MeV, whereas the binding energy is reduced to $-12.25$ MeV for $I=\frac{3}{2}$, and the $I=\frac{5}{2}$ state becomes unbound. A similar feature can also be observed in other systems. In general, the low-isospin channels tend to support deeper bound solutions, while increasing the isospin weakens the attraction and makes the formation of bound states less favorable.

       For a fixed isospin, most of the bound channels show a tendency that the binding becomes deeper in the low-$J$ states. For example, in the $\Sigma\Sigma_b^*$ system with $I=0$, the state with $J=1$ has a binding energy of $-30.58$ MeV, while the corresponding result for $J=2$ is $-21.07$ MeV. This indicates that the lower-spin channel is more favorable for binding in this case. However, this tendency is not universal. In the $\Delta\Sigma_b^*$ system with $I=\frac{1}{2}$, the deepest bound solution appears at $J=1$ rather than at $J=0$, with a binding energy of $-38.81$ MeV.  
      A similar phenomenon can also be found in hidden-charm pentaquark systems $P_c$. The spin-parity assignments of the $P_c(4440)$ and $P_c(4457)$ states are still model dependent.
      For example, Xiao $et~al$. \cite{cjx} assigned the $P_c(4440)$ and $P_c(4457)$ as $\Sigma_c\bar D^*$ molecular states with $J^P=\frac{1}{2}^{-}$ and $J^P=\frac{3}{2}^{-}$, respectively, whereas Liu $et~al$. \cite{mzl} favored the opposite assignment in the one-boson-exchange model, namely $P_c(4440)$ with $J^P=\frac{3}{2}^{-}$ and $P_c(4457)$ with $J^P=\frac{1}{2}^{-}$. In our previous study of hidden-charm pentaquark systems $P_c$ \cite{hxhajlp}, four-channel coupling results for the $\Sigma_c D^*$ channel give resonance masses of $4448.8$--$4461.6$ MeV in the $J^P=\frac{1}{2}^{-}$ sector, which are closer to the experimentally observed $P_c(4457)$. In contrast, the corresponding values in the $J^P=\frac{3}{2}^{-}$ sector are lower, around $4444.0$--$4445.7$ MeV, and are closer to the observed $P_c(4440)$. Therefore, for a fixed isospin, the relation between the energy and the total angular momentum $J$ does not show a stable regularity.
      
      We compare our results with those of previous studies.
      %For the singly bottomed dibaryon, several earlier studies focused on compact $H_b(uuddsb)$-type hexaquarks. 
      Lichtenberg $et~al$. studied the $H_b(uuddsb)$ state in a quark–diquark model and estimated its lower mass limit to be about 10 MeV below the $\Lambda\Lambda_b$ threshold, suggesting that it might be weakly bound against strong decay \cite{dbl}. However, since their method provided only a lower limit for the mass and neglected the Pauli repulsion between quarks belonging to different diquarks, the existence of a bound $H_b$ state could not be established conclusively.
        In our calculation, the $\Lambda\Lambda_b$ channel with $IJ=00$, which has the same quark content as the $H_b$ dibaryon, does not support a bound state. In contrast, the $\Sigma\Sigma_b$ and $\Sigma^{*}\Sigma_b^{*}$ channels are found to be bound, with binding energies of -43.13 and -33.68 MeV, respectively. Therefore, the present single-channel results alone do not support a bound $H_b$ state. A full coupled-channel calculation in $IJ=00$ is therefore necessary to clarify this possibility and will be pursued in our future work.
       Although the existence of the $H$-particle has not yet been conclusively established, $H$-like dibaryon states containing heavy quarks are worthy of further theoretical and experimental exploration.
       
       		\begin{figure*}[t]
	\centering
	\begin{minipage}[c]{0.04\textwidth}
		\centering
		\vspace{0.0cm}
		\rotatebox{90}{$V(S_i)\,(\mathrm{MeV})$}
	\end{minipage}
	\hspace{-0.01\textwidth}
	\begin{minipage}[c]{0.80\textwidth}
		\centering
		\includegraphics[width=\linewidth]{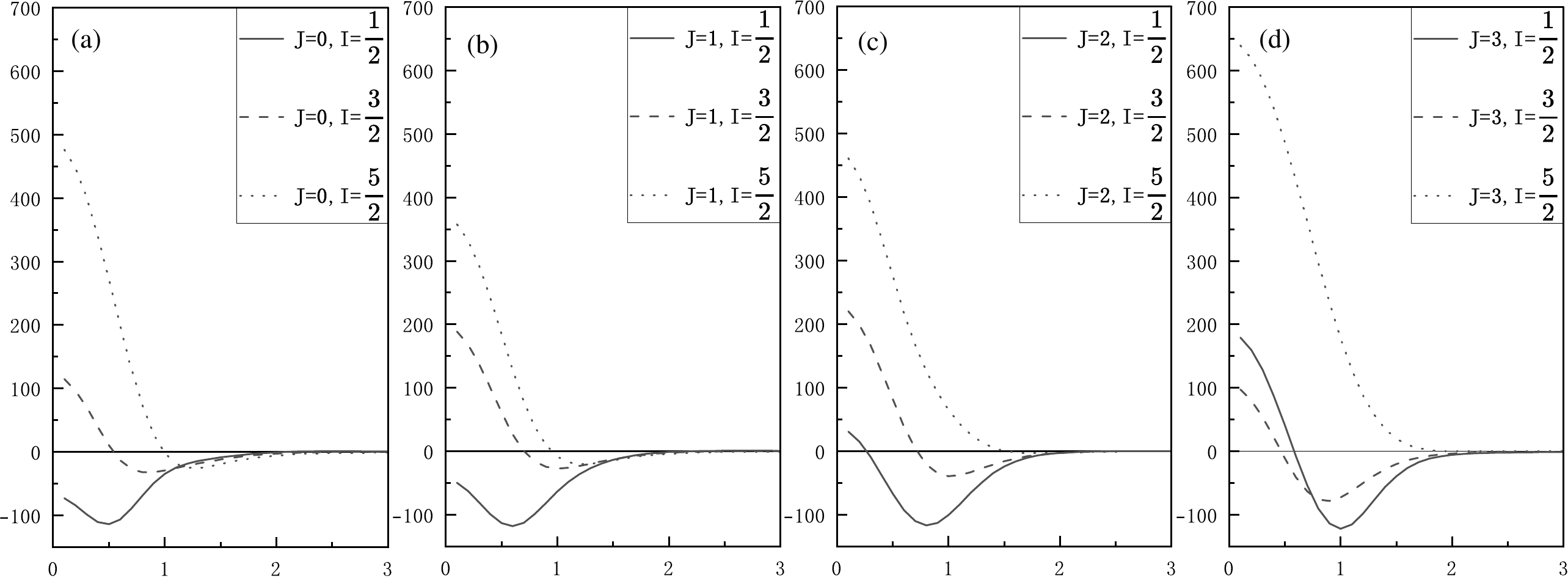}
	\end{minipage}
	\vspace{0.10cm}
	\makebox[\textwidth][c]{$S_i\,(\mathrm{fm})$}
	\caption{The effective potentials of $\Delta\Sigma_b^*$ for different $J$ and $I$.}
	\label{Figure1}
\end{figure*}

\begin{figure*}[t]
	\centering
	\begin{minipage}[c]{0.04\textwidth}
		\centering
		\vspace{0.0cm}
		\rotatebox{90}{$V(S_i)\,(\mathrm{MeV})$}
	\end{minipage}
	\hspace{-0.01\textwidth}
	\begin{minipage}[c]{0.82\textwidth}
		\centering
		\includegraphics[width=\linewidth]{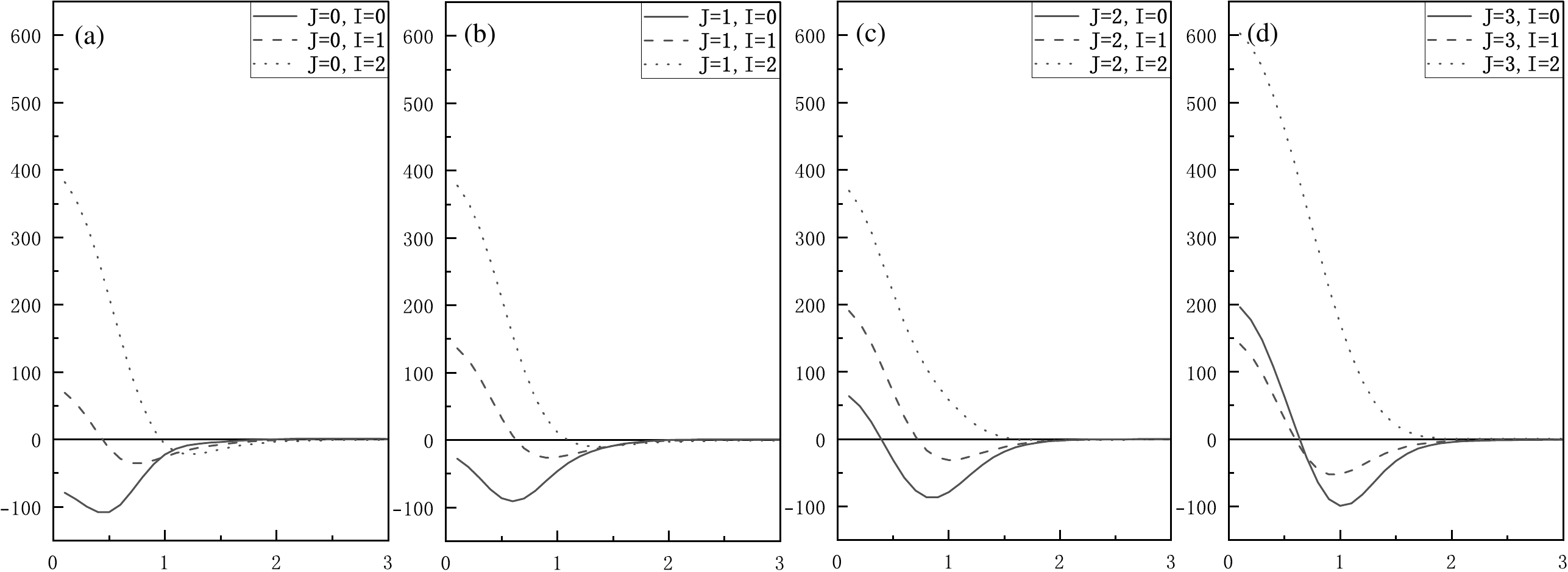}
	\end{minipage}
	\vspace{0.10cm}
	\makebox[\textwidth][c]{$S_i\,(\mathrm{fm})$}
	\caption{The effective potentials of $\Sigma^*\Sigma_b^*$ for different $J$ and $I$.}
	\label{Figure4}
\end{figure*}
      
     Leandri and Silvestre-Brac investigated heavy-flavor dibaryons within the chromomagnetic model and reported several possible bound configurations \cite{jl}.~For a given quark-flavor content, the lowest-isospin states are always the most deeply bound, in agreement with our finding that lower-isospin channels generally exhibit stronger attraction and are more favorable for binding. Among the bound states identified in their study, several bound states containing a single bottom quark are discovered, which are the $uuddsb$ states with $IJ=02$ and $IJ=03$, the $uudssb$ state with $IJ=\frac{1}{2}2$, and the $udsssb$ state with $IJ=02$. In our work, four bound states are obtained in the $qqqqsb$ quark content, corresponding to the $\Sigma\Sigma_b$ with $IJ=00,~01$, and $11$, the $\Sigma\Sigma_b^{*}$ with $IJ=01$ and $02$, the $\Sigma^*\Sigma_b$ with $IJ=01$ and $02$, the $\Sigma^*\Sigma_b^{*}$ with $IJ=00,~01,~02,~03$ and $13$.
    However, no bound solutions are obtained for the $qqqssb$ state and for the $qqsssb$ system. In addition, we also find several bound states in the $qqqqqb$ quark content, which are $\Delta\Sigma_b^*$ with $IJ=\frac{1}{2}0$, $\frac{1}{2}1$, $\frac{1}{2}2$, $\frac{1}{2}3$, and $\frac{3}{2}3$, $\Delta\Sigma_b$ with $IJ=\frac{1}{2}1$, $\frac{3}{2}1$, $\frac{1}{2}2$, and $\frac{3}{2}2$, the $N\Sigma_b^*$ with $IJ=\frac{1}{2}2$. 
    The difference between our results and those of Ref.~\cite{jl} may arise from the different models. Their calculation was performed within a simple chromomagnetic model. In contrast, the chiral quark model used in the present work not only includes the color-magnetic interaction, but also the confinement, kinetic-energy, scalar-meson-exchange, and other meson-exchange contributions. Therefore, the binding behavior obtained in our calculation results from the combined effects of several interaction terms.
    To further investigate the roles of these interaction terms, we analyze the corresponding baryon-baryon interactions in more detail to clarify the underlying binding mechanism.

%     The investigations of the $qqqqqb$ dibaryon system are still relatively few. Gerasyuta $et~al$.~\cite{smg} investigated the $qqqqqQ$ systems with one bottom quark in a relativistic approach using the dispersion relation technique. By searching for the poles of the six-quark amplitudes, they obtained the masses of bottomed dibaryon states with isospins $\frac{1}{2},~\frac{3}{2},~\frac{5}{2}$ and predicted possible bottomed dibaryons.

     %we calculate the effective potentials of several representative channels and analyze the contributions from different interaction terms.

%To investigate possible bound states, we conducted systematic bound state calculations. We calculated the energies of each single channel. Meanwhile, we also consider the effect of channel coupling and perform coupled channel calculations, and the computational results are listed in Tables \ref{tab1}-\ref{tab3}, which are expressed in the natural unit system ($\hbar$ = $c$ = 1).

\subsection{The analysis of interactions}

	In present work, the effective potentials are used to study the interaction between two baryons, which is defined as the following form:
	\begin{align}
		V\left(S_i\right)=E\left(S_i\right)-E\left(\infty\right),
	\end{align}
	where $S_i$ denotes the separation distance between the two baryon clusters, and the energy $E\left(S_i\right)$ is explicitly given by:
	\begin{equation}
		E(S_i) = \frac{\langle \Psi_{6q}(S_i) \, | \, H \, | \, \Psi_{6q}(S_i) \rangle}{\langle \Psi_{6q}(S_i) | \Psi_{6q}(S_i) \rangle},
	\end{equation}
	$\Psi_{6q}(S_i)$ is defined as the wave function of a given dibaryon state, with $\langle\Psi_{6q}(S_i)\,|\,H\,|\,\Psi_{6q}(S_i)\rangle$ and $\langle \Psi_{6q}(S_i) | \Psi_{6q}(S_i) \rangle$ being the Hamiltonian matrix element and the norm of the state, respectively. 
	
\begin{figure}[H]
	\centering
	\begin{subfigure}{1.0\linewidth}
		\centering
		\begin{minipage}[c]{0.05\linewidth}
			\centering
			\rotatebox{90}{$V(S_i)\,(\mathrm{MeV})$}
		\end{minipage}
		\hspace{-0.01\linewidth}
		\begin{minipage}[c]{0.86\linewidth}
			\centering
			\includegraphics[width=\linewidth]{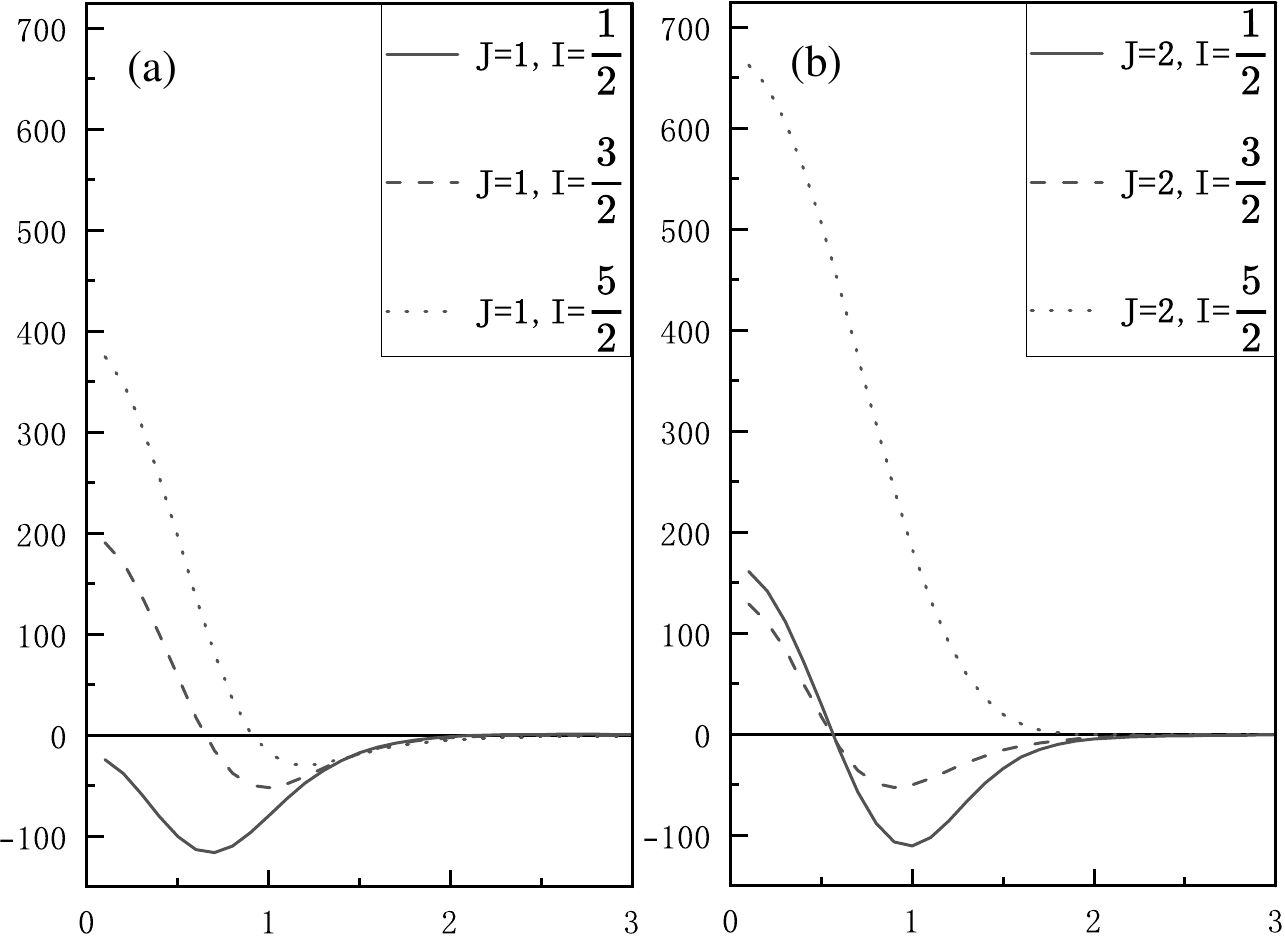}
		\end{minipage}
	\end{subfigure}
	\vspace{0.3cm}
	{\centering $S_i\,(\mathrm{fm})$\par}
	\caption{The effective potentials of $\Delta\Sigma_b$ for different $J$ and $I$.}
	\label{Figure2}
\end{figure}

\begin{figure}[H]
	\centering
	\begin{subfigure}{1.0\linewidth}
		\centering
		\begin{minipage}[c]{0.05\linewidth}
			\centering
			\rotatebox{90}{$V(S_i)\,(\mathrm{MeV})$}
		\end{minipage}
		\hspace{-0.01\linewidth}
		\begin{minipage}[c]{0.86\linewidth}
			\centering
			\includegraphics[width=\linewidth]{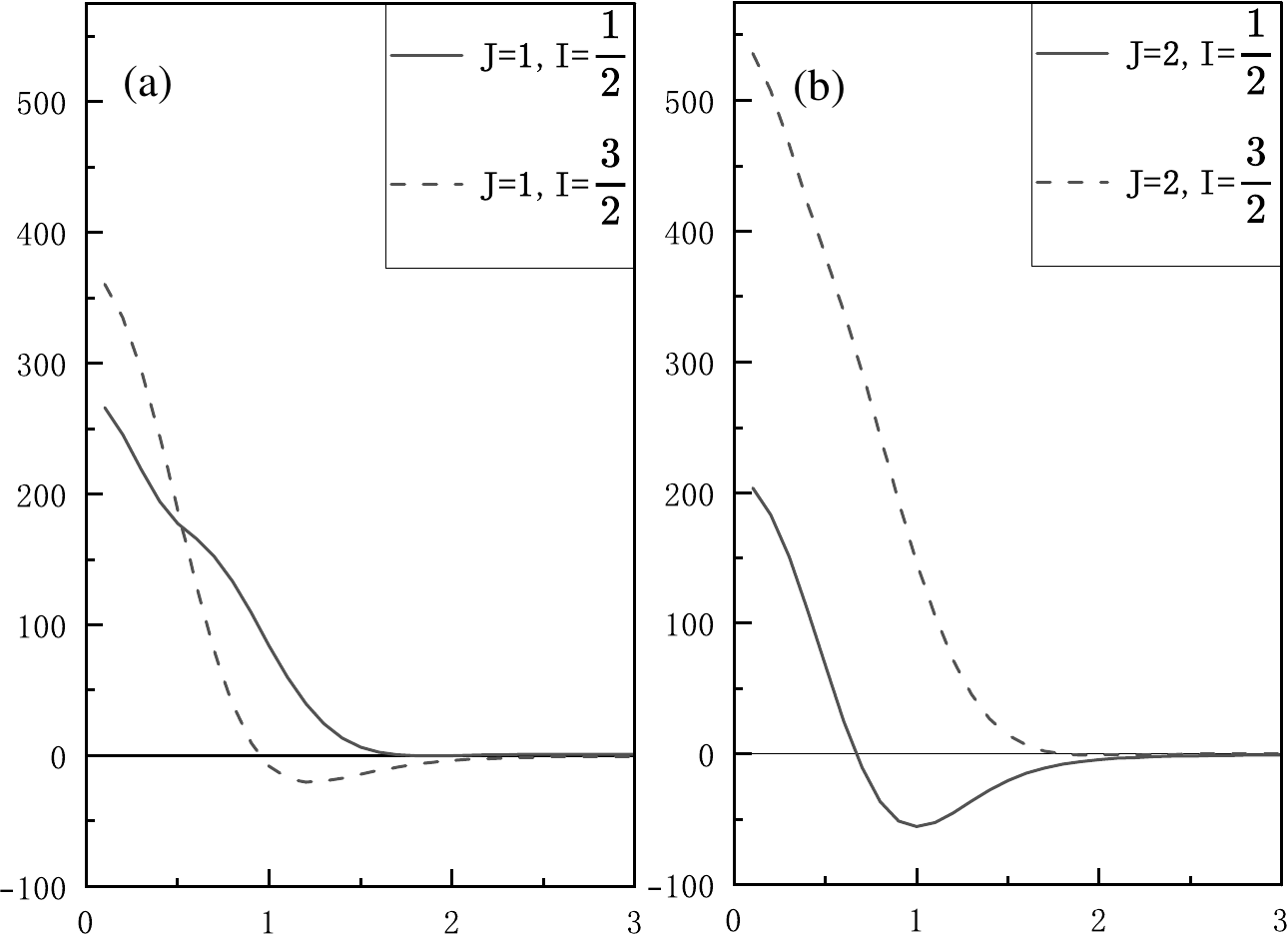}
		\end{minipage}
	\end{subfigure}
	\vspace{0.3cm}
	{\centering $S_i\,(\mathrm{fm})$\par}
	\caption{The effective potentials of $N\Sigma_b^*$ for different $J$ and $I$.}
	\label{Figure3}
\end{figure}

\begin{figure}[H]
	\centering
	\begin{subfigure}{1.0\linewidth}
		\centering
		\begin{minipage}[c]{0.05\linewidth}
			\centering
			\rotatebox{90}{$V(S_i)\,(\mathrm{MeV})$}
		\end{minipage}
		\hspace{-0.01\linewidth}
		\begin{minipage}[c]{0.86\linewidth}
			\centering
			\includegraphics[width=\linewidth]{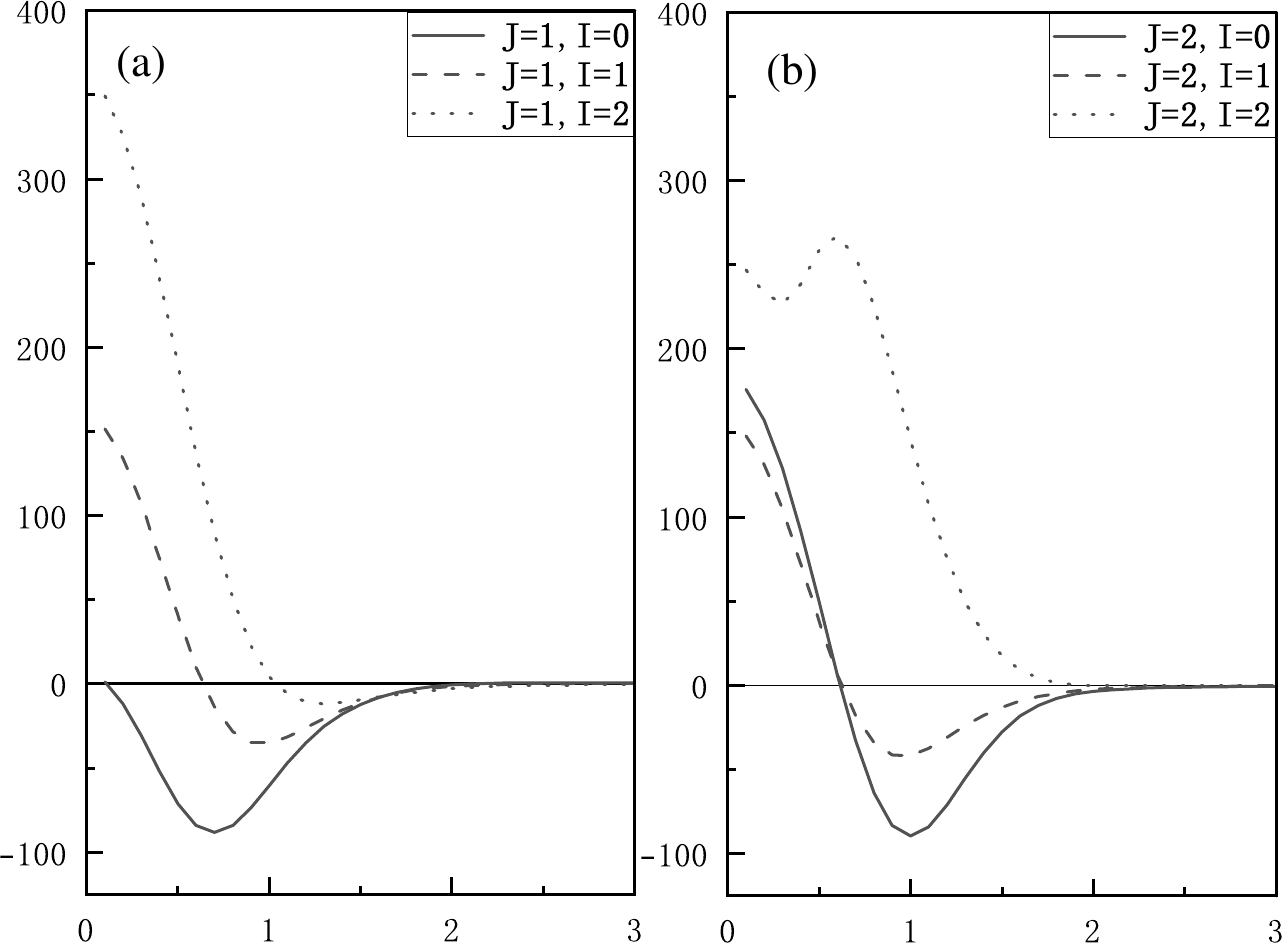}
		\end{minipage}
	\end{subfigure}
	\vspace{0.3cm}
	{\centering $S_i\,(\mathrm{fm})$\par}
	\caption{The effective potentials of $\Sigma^*\Sigma_b$ for different $J$ and $I$.}
	\label{Figure5}
\end{figure}

\begin{figure}[H]
	\centering
	\begin{subfigure}{1.0\linewidth}
		\centering
		\begin{minipage}[c]{0.05\linewidth}
			\centering
			\rotatebox{90}{$V(S_i)\,(\mathrm{MeV})$}
		\end{minipage}
		\hspace{-0.01\linewidth}
		\begin{minipage}[c]{0.86\linewidth}
			\centering
			\includegraphics[width=\linewidth]{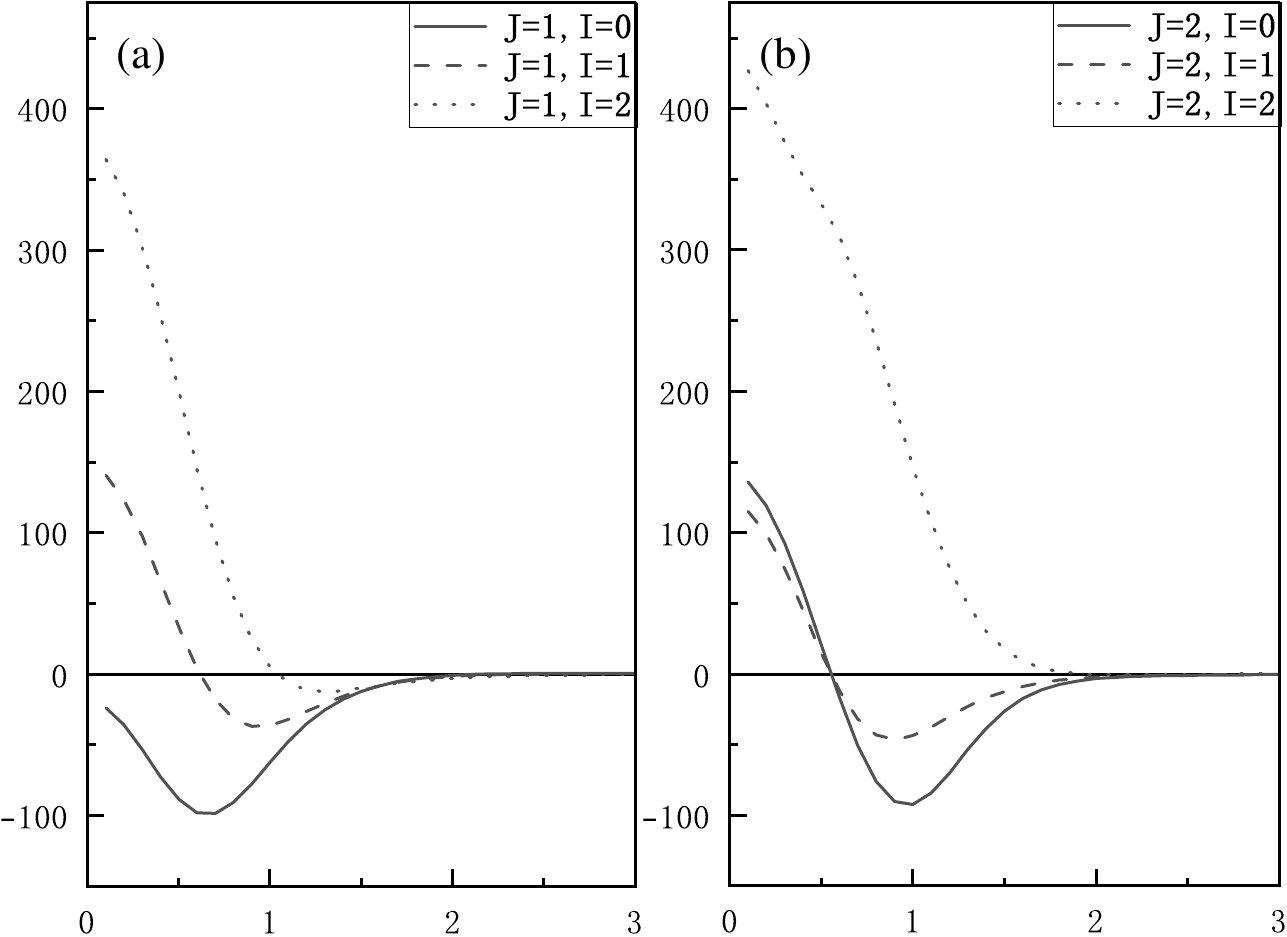}
		\end{minipage}
	\end{subfigure}
	\vspace{0.3cm}
	{\centering $S_i\,(\mathrm{fm})$\par}
	\caption{The effective potentials of $\Sigma\Sigma_b^*$ for different $J$ and $I$.}
	\label{Figure6}
\end{figure}

\begin{figure}[H]
	\centering
	\begin{subfigure}{1.0\linewidth}
		\centering
		\begin{minipage}[c]{0.05\linewidth}
			\centering
			\rotatebox{90}{$V(S_i)\,(\mathrm{MeV})$}
		\end{minipage}
		\hspace{-0.01\linewidth}
		\begin{minipage}[c]{0.86\linewidth}
			\centering
			\includegraphics[width=\linewidth]{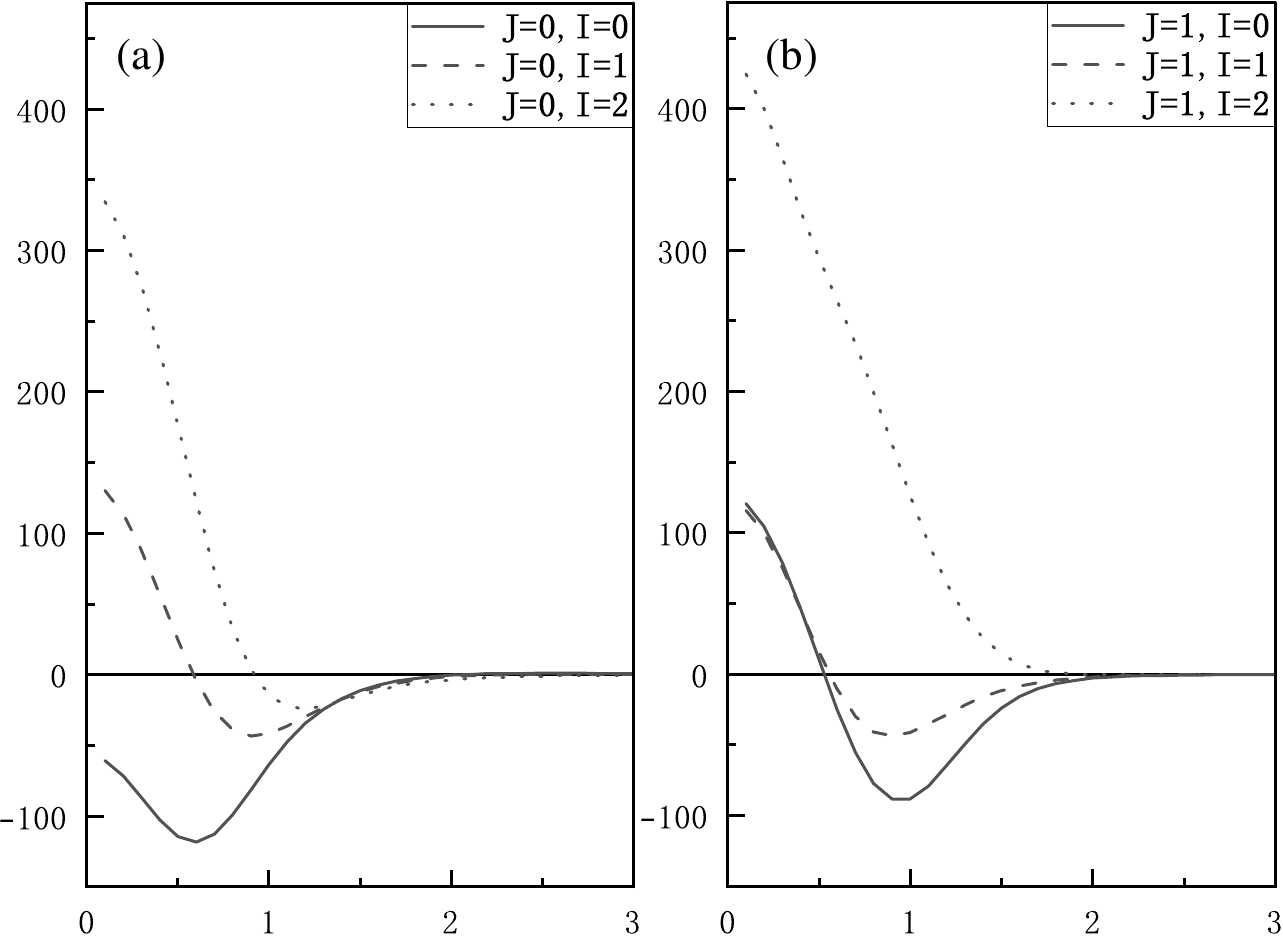}
		\end{minipage}
	\end{subfigure}
	\vspace{0.3cm}
	{\centering $S_i\,(\mathrm{fm})$\par}
	\caption{The effective potentials of $\Sigma\Sigma_b$ for different $J$ and $I$.}
	\label{Figure7}
\end{figure}

	 We then discuss the interactions, concentrating on the most interesting channels, such as $\Delta\Sigma_b^*$, $\Delta\Sigma_b$, $\Sigma\Sigma_b$, $\Sigma\Sigma^*_b$, $\Sigma^*\Sigma_b$, and $\Sigma^*\Sigma^*_b$.
	The effective potentials for systems with different quantum numbers are presented in Figs.~\ref{Figure1}-\ref{Figure7}. Although most of these effective potentials contain attractive and repulsive regions, the presence of a negative minimum indicates an attractive interaction between the two hadrons, which may allow the system to form a bound state. To save space, we only analyze the effective potentia of the $\Delta\Sigma_b^*$ and $\Delta\Sigma_b$ systems. the other states display similar features.

      Fig.~\ref{Figure1} shows the effective potential of $\Delta\Sigma_b^*$ system. We initially analyze the behavior of the effective potential for fixed $J$ with different $I$. For $J=0$, as shown in Fig.~\ref{Figure1}(a), the depth of the attractive effective potential exceeds 100 MeV at a distance of around 0.5 fm between the two clusters, which means that it is likely to form bound state. In contrast, the potential for $I=\frac{3}{2}$ is considerably shallower, while for the $I=\frac{5}{2}$ potential is the shallowest among the three. This suggeats that the effective potential of the $I=\frac{3}{2}$ and $I=\frac{5}{2}$ exhibit weak attractive, so bound states are unlikely to form in these systems. This is consistent with the results of our binding energy calculations. A similar behavior is observed for $J=1,~J=2,~J=3$, as shown in Fig.~\ref{Figure1}(b)-(d). We therefore do not discuss these panels in detail.

     We then investgate the effective potentials of $\Delta\Sigma_b^*$ sysytem with different~$J$~at~fixed~$I$. For $I=\frac{1}{2}$ state, as shown in from Fig.~\ref{Figure1}(a)-Fig.~\ref{Figure1}(d), the effective potentials for all four $J$ values exhibit attractive potential in the intermediate distance region. The corresponding potentials all exceed 100 MeV in depth and are comparable in magnitude, indicating that these states are likely to form bound states with comparable binding energies. This observation supports our bound energy calculations in Table~\ref{bound energy1}, approximately 35 MeV.
     For the $I=\frac{3}{2}$ systems, the feature on $J$ becomes more evident. The attractive potentials in the $J=0$, $J=1$, and $J=2$ states are too shallow to support bound states. In contrast, the $J = 3$ state develops a deeper attractive effective potential and produces a bound state with a binding energy of $-12.25$ MeV (see Table~\ref{bound energy1}).
     For the highest-isospin case $I=\frac{5}{2}$, the potentials remain predominantly repulsive or only very weakly attractive for all four $J$ values. No sufficiently deep attractive potential appears in these channels, and the corresponding bound state calculations give unbound solutions.

     Fig.~\ref{Figure2} shows the effective potential of $\Delta\Sigma_b$ system. Different from the $\Delta\Sigma_b^*$ case,  only the $J=1$ and $J=2$ systems are allowed for this configuration.
For the fixed $J=1$ systems, the $I=\frac{1}{2}$ potential exhibits 
a clear attractive potential at a distance of around 0.6 fm, while the attraction is strongly reduced when the isospin increases to $I=\frac{3}{2}$. In the $I=\frac{5}{2}$ system, the potential becomes predominantly repulsive and no bound state is formed. This behavior is consistent with 
the bound state calculation, where the $I=\frac{1}{2}$ state is deeply bound with a binding energy of $-38.43$ MeV, the $I=\frac{3}{2}$ state is only marginally bound by $-0.31$ MeV, and the $I=\frac{5}{2}$ state is unbound. A similar characteristic is also found in the fixed $J=2$ systems.

    Nevertheless, the same isospin hierarchy is clearly visible. For both $J=1$ and $J=2$, the $I=\frac{1}{2}$ potentials show distinct attraction, indicating that the interaction is sufficiently strong to generate bound solutions. This is consistent with the bound state exhibited in Table~\ref{bound energy1}, where the corresponding binding energies are $-38.43$ MeV and $-28.68$ MeV, respectively. 
When the isospin is increased to $I=\frac{3}{2}$, the attraction is strongly reduced. The potential becomes very shallow and lies close to the threshold, leading to bound states with marginal binding energies of $-0.31$ MeV for $J=1$ and $-1.55$ MeV for $J=2$. For the highest-isospin $I=\frac{5}{2}$ system, the interaction is dominated by repulsion at short distances and no bound state is formed. 


%It can be seen that, for a fixed total angular momentum $J$, the attractive interaction becomes weaker as the isospin $I$ increases.
%On the other hand, for a fixed isospin, the effective potentials with different $J$ values provides a explanation for the deepest bound $\Delta\Sigma_b^*$ state does not necessarily occur in the lowest-$J$ channel.


  From the above discussion, we can conclude that the depth and shape of the effective potential are not governed solely by $J$ or $I$, but rather by the combined effects of isospin-spin coupling.  

\begin{figure}[H]
	\centering
	\begin{subfigure}{1.0\linewidth}
		\centering
		\begin{minipage}[c]{0.02\linewidth}
			\centering
			\rotatebox{90}{$V(S_i)\,(\mathrm{MeV})$}
		\end{minipage}
		\hspace{-0.01\linewidth}
		\begin{minipage}[c]{0.9678\linewidth}
			\centering
			\includegraphics[width=\linewidth]{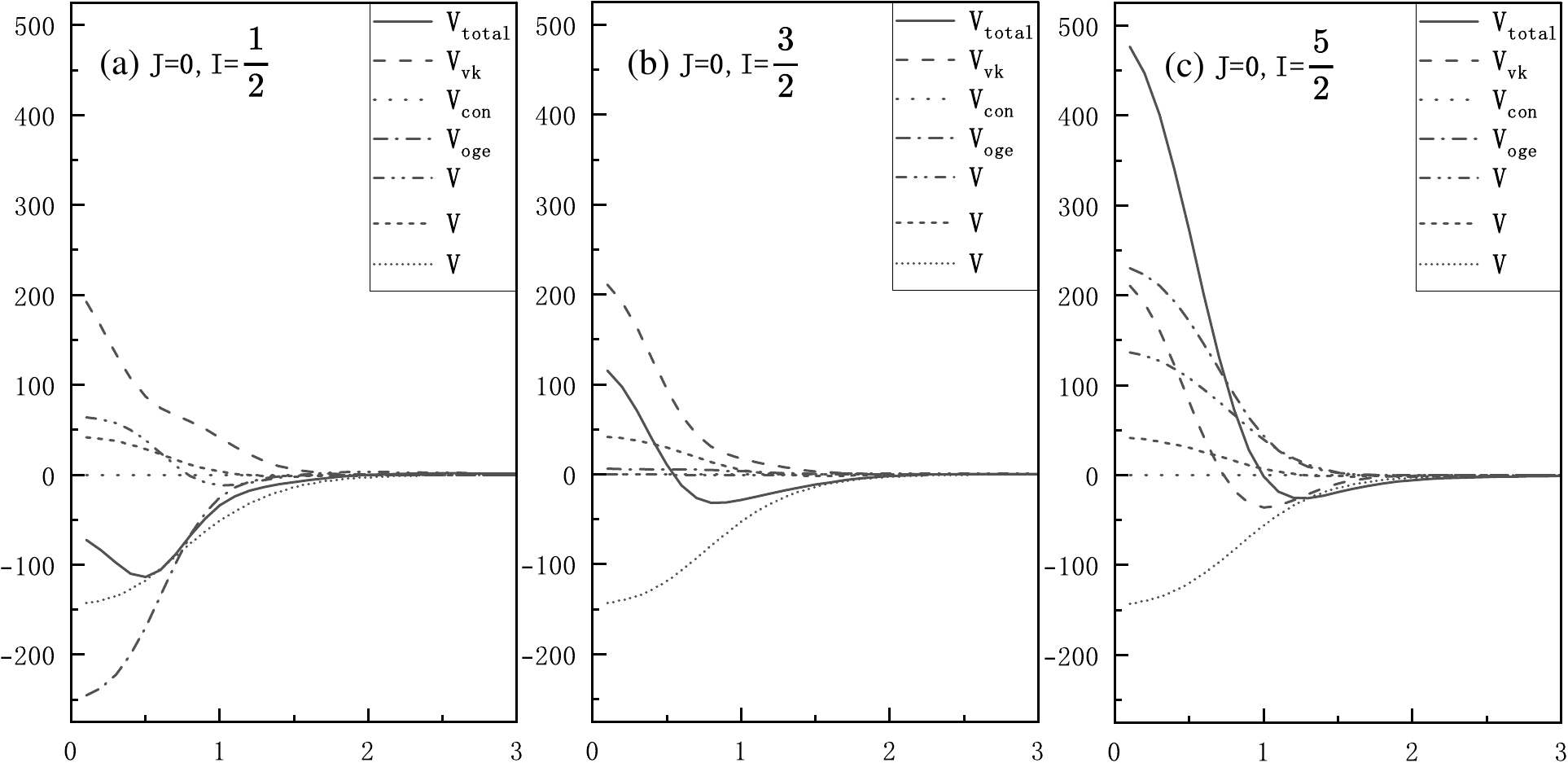}
		\end{minipage}
	\end{subfigure}
	\vspace{0.1cm}
	{\centering $S_i\,(\mathrm{fm})$\par}
	\caption{Contributions of the different terms of the interaction to the $J=0$ $\Delta\Sigma_b^*$ potential.}
	\label{Figure8}
\end{figure}

\begin{figure}[H]
	\centering
	\begin{subfigure}{1.0\linewidth}
		\centering
		\begin{minipage}[c]{0.02\linewidth}
			\centering
			\rotatebox{90}{$V(S_i)\,(\mathrm{MeV})$}
		\end{minipage}
		\hspace{-0.01\linewidth}
		\begin{minipage}[c]{0.9678\linewidth}
			\centering
			\includegraphics[width=\linewidth]{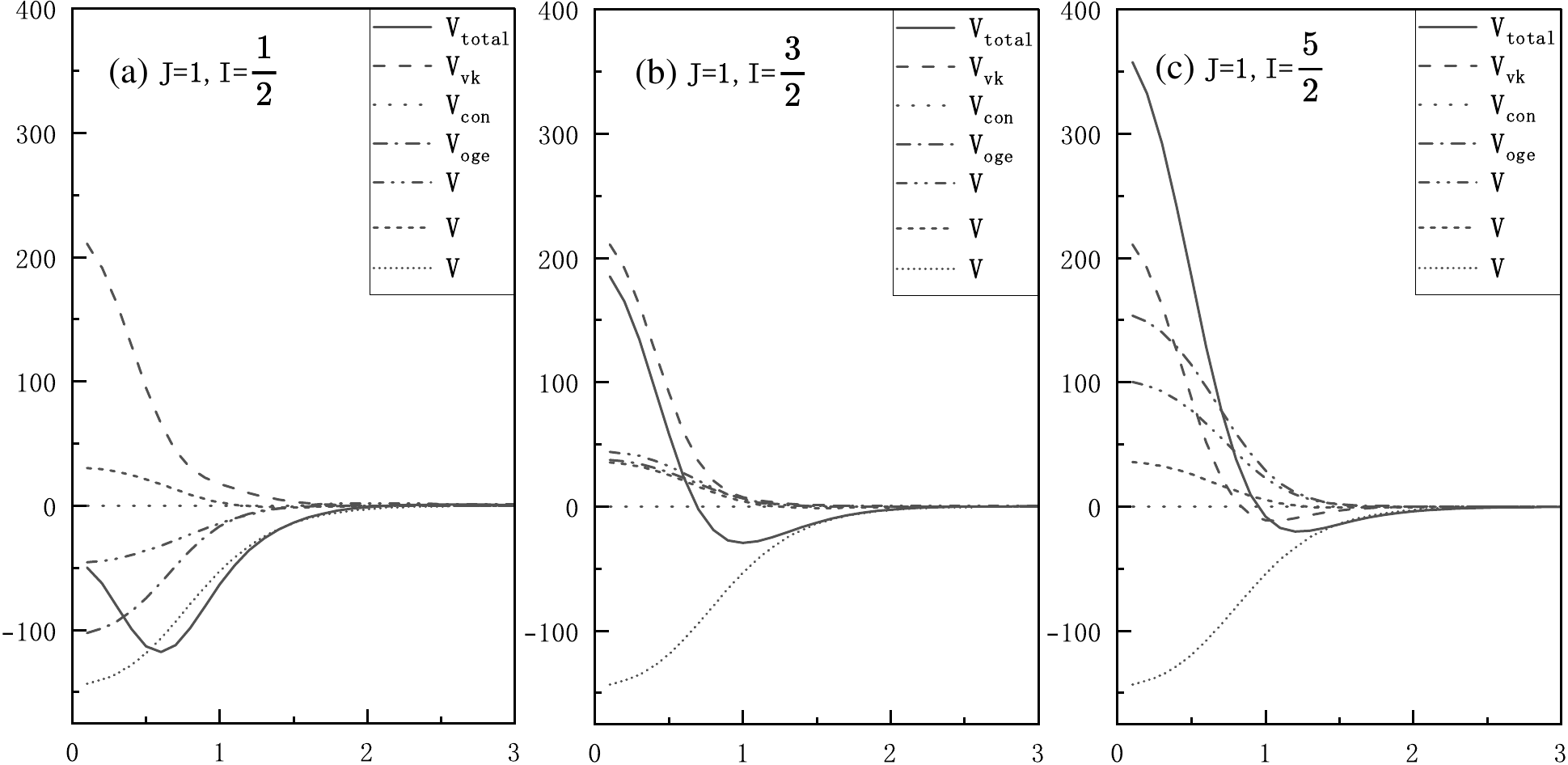}
		\end{minipage}
	\end{subfigure}
	\vspace{0.1cm}
	{\centering $S_i\,(\mathrm{fm})$\par}
	\caption{Contributions of the different terms of the interaction to $J=1$ $\Delta\Sigma_b^*$ potential.}
	\label{Figure9}
\end{figure}

\begin{figure}[H]
	\centering
	\begin{subfigure}{1.0\linewidth}
		\centering
		\begin{minipage}[c]{0.02\linewidth}
			\centering
			\rotatebox{90}{$V(S_i)\,(\mathrm{MeV})$}
		\end{minipage}
		\hspace{-0.01\linewidth}
		\begin{minipage}[c]{0.9678\linewidth}
			\centering
			\includegraphics[width=\linewidth]{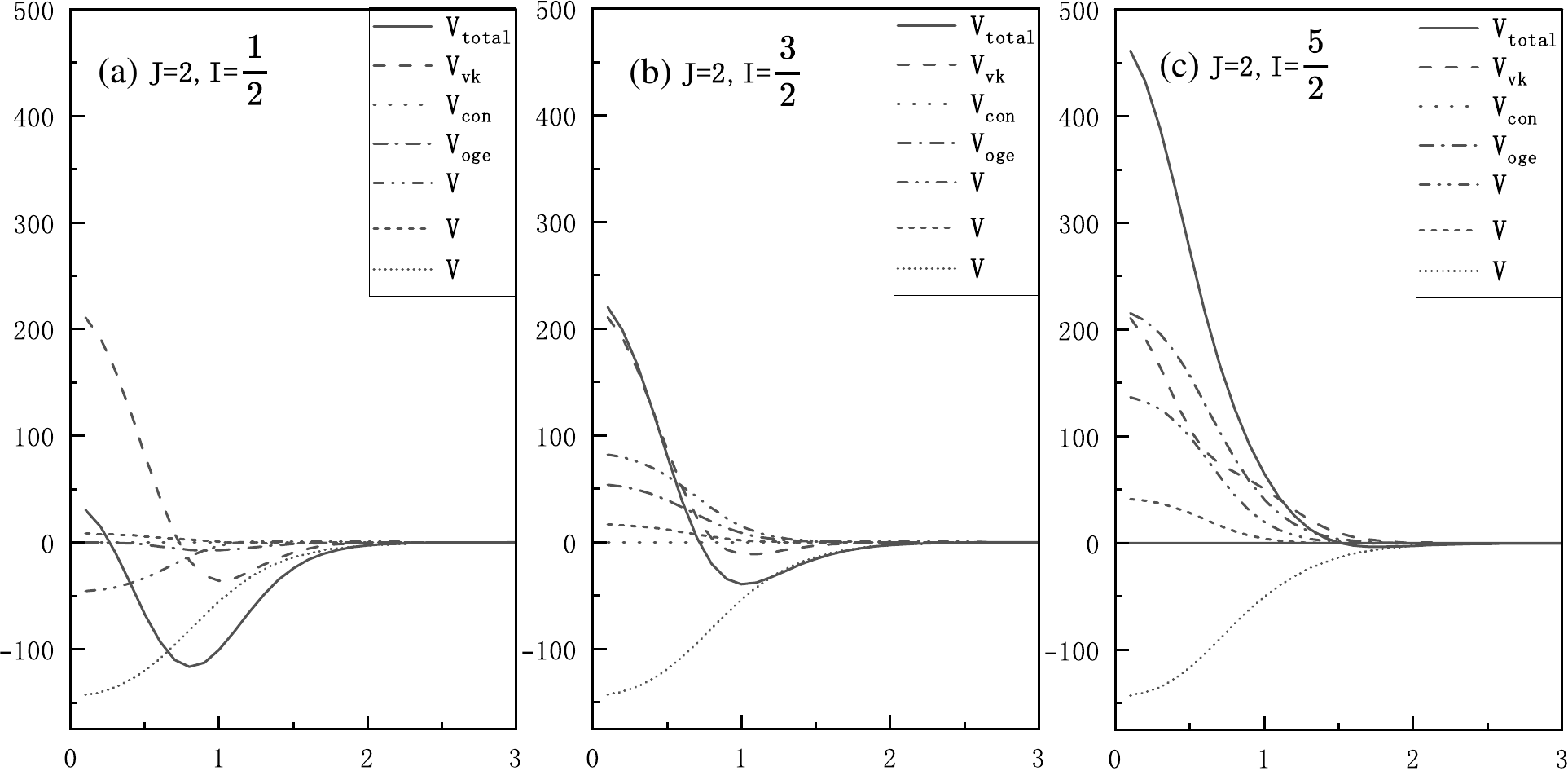}
		\end{minipage}
	\end{subfigure}
	\vspace{0.1cm}
	{\centering $S_i\,(\mathrm{fm})$\par}
	\caption{Contributions of the different terms of the interaction to $J=2$ $\Delta\Sigma_b^*$ potential.}
	\label{Figure10}
\end{figure}

\begin{figure}[H]
	\centering
	\begin{subfigure}{1.0\linewidth}
		\centering
		\begin{minipage}[c]{0.02\linewidth}
			\centering
			\rotatebox{90}{$V(S_i)\,(\mathrm{MeV})$}
		\end{minipage}
		\hspace{-0.01\linewidth}
		\begin{minipage}[c]{0.9678\linewidth}
			\centering
			\includegraphics[width=\linewidth]{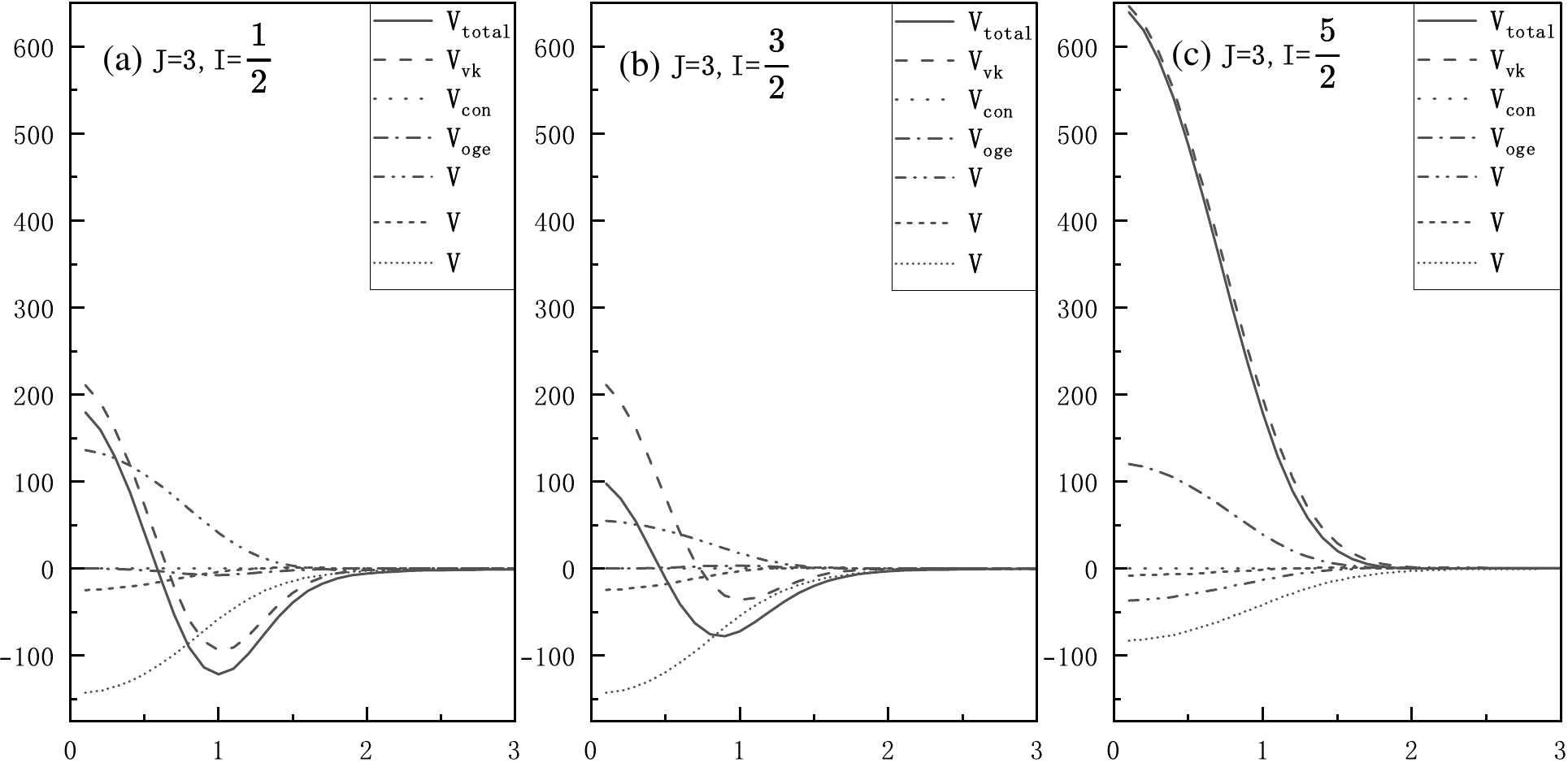}
		\end{minipage}
	\end{subfigure}
	\vspace{0.1cm}
	{\centering $S_i\,(\mathrm{fm})$\par}
	\caption{Contributions of the different terms of the interaction to $J=3$ $\Delta\Sigma_b^*$ potential.}
	\label{Figure11}
\end{figure}

  To identify the origin of the attractive interaction, we analyze the individual contributions of the interaction terms. In our quark model, the interaction consists of the kinetic energy ($V_{vk}$), the confinement ($V_{con}$), the one-gluon exchange ($V_{oge}$), and the meson exchange. Since different terms may dominate in different distance regions, such a decomposition is helpful for identifying the interaction mechanism.
    We have analyze the seven states listed in Tables \ref{bound energy1} and \ref{bound energy2}, and found that the mechanisms are similar. Therefore, to avoid unnecessary repetition and to save space, we only present the results of $\Delta\Sigma_b^*$ system with $J=0$, $1$, $2$, and $3$, which are displayed in Figs. \ref{Figure8}-\ref{Figure11}, respectively. Since the contributions of the $B$, $B_{s}$ and $\eta_{b}$ meson exchange potentials are found to be insignificant, they are not discussed in detail in the following analysis.

     We first focus on the $J=0$ systems shown in Fig.~\ref{Figure8}. For the $I=\frac{1}{2}$ state, the total effective potential exhibits a clear attractive interaction. One can see that there is no contribution of color-confinement interaction between two color-singlet clusters. The attraction mainly originates from the scalar $\sigma$ meson exchange in the intermediate distances and the one-gluon exchange interaction at short distances. In contrast, the kinetic term gives a sizable repulsive effect at short distances. The attractive contributions overcome the repulsive components and consequently lead to a negative total potential in the intermediate-range. 
     For the $I=\frac{3}{2}$ state, the scalar $\sigma$ meson exchange still provides an attractive contribution in the intermediate distance region, and the other terms do not change significantly compared with the $I=\frac{1}{2}$ state. However, the one-gluon exchange interaction becomes weakly repulsive. Although this repulsion is not very strong,  the sum of the repulsive components partially cancels the attraction generated by the $\sigma$ meson exchange and other attractive terms. As a result, the total effective potential becomes much shallower than that in the $I=\frac{1}{2}$ state, which explains why the attractive interaction is strongly weakened when the isospin increases from $I=\frac{1}{2}$ to $I=\frac{3}{2}$.
    For the $I=\frac{5}{2}$ state, the $\sigma$ meson exchange remains attractive, and the kinetic-energy term also gives a weak attractive contribution. Nevertheless, the one-gluon exchange interaction becomes strongly repulsive over the short-range region. This repulsive contribution largely cancels the attraction from the $\sigma$ meson exchange and the kinetic term. Consequently, the total effective potential only shows a very weak attraction and is not sufficient to support a bound state.
    
  Since the behaviors observed for $J=1,~2$ systems are similar to  $J=0$. We will not discuss these systems in detail here.
    For the $J=3$ $\Delta\Sigma_b^*$ system, the decomposed potentials are shown in Fig.~\ref{Figure11}. In the $I=\frac{1}{2}$ state, the total effective potential exhibits attractive behavior. 
The scalar $\sigma$ meson exchange still provides the dominant attraction in the intermediate distance region. In addition, the kinetic energy term also gives an attractive contribution in this channel, which further deepens the total potential. The one-gluon exchange contribution is relatively weak and does not destroy the attractive behavior. 
As a result, the combined attraction is strong enough to generate a bound state, consistent with the binding energy listed in Table \ref{bound energy1}.

    For the $I=\frac{3}{2}$ state, the $\sigma$ meson exchange remains attractive, but the attractive contribution from the kinetic energy term is reduced compared with the $I=\frac{1}{2}$ case. Meanwhile, the one-gluon exchange interaction gives a repulsive contribution in the short-range region. The cancellation between the remaining $\sigma$ meson attraction and the repulsive components makes the total effective potential much shallower. 
Therefore, although this channel still supports a bound solution, the binding becomes much weaker than that in the $I=\frac{1}{2}$ state.
    For the highest-isospin $I=\frac{5}{2}$ state, the reduction of attraction becomes more evident. Although the scalar $\sigma$ meson exchange still provides an attractive contribution, it is no longer sufficient to overcome the short-range repulsion. In this channel, the light-quark flavor configuration is highly symmetric, and the $J=3$ state also corresponds to a highly symmetric spin configuration. Together with the $S$-wave spatial symmetry, the Pauli exclusion principle imposes a stronger constraint on the allowed six-quark configuration. This Pauli blocking effect enhances the short-range repulsion generated by quark exchange and strongly suppresses the overlap between the two baryon clusters. Consequently, the total effective potential becomes strongly repulsive, and no bound state is obtained in the $I=\frac{5}{2}$ state.
   

    %Given that the contributions of the other particles $\Delta\Sigma_{b}$ and $\Sigma^*\Sigma_{b}^*$, as illustrated in the Figs \ref{Figure12}-\ref{Figure13}, are anamologous. They are not discussed further here
    To further investigate the origin of different contributions to the effective potentials, we calculate the matrix elements in the spin-flavor-color space. 
These matrix elements serve as spin-flavor-color coefficients in the direct and exchange terms, and determine the relative contribution of each quark pair as well as its dependence on different quantum number channels. 
Therefore, they are useful for clarifying the microscopic origin of the effective potentials. 
The relevant matrix elements are defined as follows:
 
% To provide more information for the six-quark system calculation, the matrix elements in the spin-flavor-color space are listed in Tables \ref{matrix1}. To save space, only the result for the representative $\Delta\Sigma_b^*$ channel is presented. These matrix elements, serving as spin-flavor-color coefficients in the direct and exchange terms, determine both the quark-pair contribution to each term and its variation with quantum number channels.
 %Therefore, they are useful for clarifying the microscopic origin of the effective potentials. The definitions of the matrix elements are given as follows:
 \begin{align}
N_D &= \langle \psi | \psi \rangle, \\
N_E &= \langle \psi | \mathcal{A}' | \psi \rangle, \\
\langle H_{ij}^{c} \rangle_D
&= \langle \psi | \boldsymbol{\lambda}_i^{c} \cdot \boldsymbol{\lambda}_j^{c} | \psi \rangle, \\
\langle H_{ij}^{c} \rangle_E
&= \langle \psi | \boldsymbol{\lambda}_i^{c} \cdot \boldsymbol{\lambda}_j^{c} \mathcal{A}' | \psi \rangle, \\
\langle H_{ij}^{cs} \rangle_D
&= \langle \psi | \boldsymbol{\lambda}_i^{c} \cdot \boldsymbol{\lambda}_j^{c}
\boldsymbol{\sigma}_i \cdot \boldsymbol{\sigma}_j | \psi \rangle, \\
\langle H_{ij}^{cs} \rangle_E
&= \langle \psi | \boldsymbol{\lambda}_i^{c} \cdot \boldsymbol{\lambda}_j^{c}
\boldsymbol{\sigma}_i \cdot \boldsymbol{\sigma}_j \mathcal{A}' | \psi \rangle.  
\end{align}
	\begin{table*}[t]
	\centering
	\renewcommand{\arraystretch}{1.35}
	\caption{The matrix elements in spin-flavor-color-isospin spaces of the color-singlet $\Delta \Sigma^*_b$ channel with $J=0$ and different isospins.}
	\label{matrix1}
\begin{tabular}{c cccc c cccc c cccc}
		\hline
		\hline
		\multicolumn{5}{c}{$IJ=\frac{1}{2}0$}
		& \multicolumn{5}{c}{$IJ=\frac{3}{2}0$}
		& \multicolumn{5}{c}{$IJ=\frac{5}{2}0$} \\
		\hline
		$ij$ & ${\left\langle H_{ij}^c \right\rangle}_{D}$ & ${\left\langle H_{ij}^c \right\rangle}_{E}$ 
		& ${\left\langle H_{ij}^{cs} \right\rangle}_{D}$ & ${\left\langle H_{ij}^{cs} \right\rangle}_{E}$
		&
		~~~ & ${\left\langle H_{ij}^c \right\rangle}_{D}$ & ${\left\langle H_{ij}^c \right\rangle}_{E}$ 
		& ${\left\langle H_{ij}^{cs} \right\rangle}_{D}$ & ${\left\langle H_{ij}^{cs} \right\rangle}_{E}$
		&
		~~~ & ${\left\langle H_{ij}^c \right\rangle}_{D}$ & ${\left\langle H_{ij}^c \right\rangle}_{E}$ 
		& ${\left\langle H_{ij}^{cs} \right\rangle}_{D}$ & ${\left\langle H_{ij}^{cs} \right\rangle}_{E}$ \\
		\hline
		12 & $-8/3$ & $-8/9$  & $-8/3$ & $-8/9$ 
		&  & $-8/3$ & $-8/27$ & $-8/3$ & $-8/27$
		&  & $-8/3$ & $8/9$   & $-8/3$ & $8/9$ \\

		13 & $-8/3$ & $-8/9$  & $-8/3$ & $-8/9$
		&  & $-8/3$ & $-8/27$ & $-8/3$ & $-8/27$
		&  & $-8/3$ & $8/9$   & $-8/3$ & $8/9$ \\

		14 & $0$    & $4/9$   & $0$    & $-4/3$
		&  & $0$    & $4/27$  & $0$    & $-4/9$
		&  & $0$    & $-4/9$  & $0$    & $4/3$ \\

		15 & $0$    & $4/9$   & $0$    & $-4/3$
		&  & $0$    & $4/27$  & $0$    & $-4/9$
		&  & $0$    & $-4/9$  & $0$    & $4/3$ \\

		16 & $0$    & $-8/9$  & $0$    & $-8/9$
		&  & $0$    & $-8/27$ & $0$    & $-8/27$
		&  & $0$    & $8/9$   & $0$    & $8/9$ \\

		23 & $-8/3$ & $-8/9$  & $-8/3$ & $-8/9$
		&  & $-8/3$ & $-8/27$ & $-8/3$ & $-8/27$
		&  & $-8/3$ & $8/9$   & $-8/3$ & $8/9$ \\

		24 & $0$    & $4/9$   & $0$    & $-4/3$
		&  & $0$    & $4/27$  & $0$    & $-4/9$
		&  & $0$    & $-4/9$  & $0$    & $4/3$ \\

		25 & $0$    & $4/9$   & $0$    & $-4/3$
		&  & $0$    & $4/27$  & $0$    & $-4/9$
		&  & $0$    & $-4/9$  & $0$    & $4/3$ \\

		26 & $0$    & $-8/9$  & $0$    & $-8/9$
		&  & $0$    & $-8/27$ & $0$    & $-8/27$
		&  & $0$    & $8/9$   & $0$    & $8/9$ \\

		34 & $0$    & $-8/9$  & $0$    & $-8/9$
		&  & $0$    & $-8/27$ & $0$    & $-8/27$
		&  & $0$    & $8/9$   & $0$    & $8/9$ \\

		35 & $0$    & $-8/9$  & $0$    & $-8/9$
		&  & $0$    & $-8/27$ & $0$    & $-8/27$
		&  & $0$    & $8/9$   & $0$    & $8/9$ \\

		36 & $0$    & $16/9$  & $0$    & $-112/9$
		&  & $0$    & $16/27$ & $0$    & $-112/27$
		&  & $0$    & $-16/9$ & $0$    & $112/9$ \\

		45 & $-8/3$ & $-8/9$  & $-8/3$ & $-8/9$
		&  & $-8/3$ & $-8/27$ & $-8/3$ & $-8/27$
		&  & $-8/3$ & $8/9$   & $-8/3$ & $8/9$ \\

		46 & $-8/3$ & $-8/9$  & $-8/3$ & $-8/9$
		&  & $-8/3$ & $-8/27$ & $-8/3$ & $-8/27$
		&  & $-8/3$ & $8/9$   & $-8/3$ & $8/9$ \\

		56 & $-8/3$ & $-8/9$  & $-8/3$ & $-8/9$
		&  & $-8/3$ & $-8/27$ & $-8/3$ & $-8/27$
		&  & $-8/3$ & $8/9$   & $-8/3$ & $8/9$ \\
		\hline
		\hline
	\end{tabular}
\end{table*}

\begin{table*}[htbp]
\centering
\footnotesize
\setlength{\tabcolsep}{3pt}
\renewcommand{\arraystretch}{1.35}
\caption{The matrix elements in the spin-flavor-color-isospin space of the color-singlet $\Delta \Sigma^*_{b}$ channel with $J=1$ and different isospins.}
\label{matrix2}
\begin{tabular}{c cccc c cccc c cccc}
\hline
\hline
\multicolumn{5}{c}{$IJ=\frac{1}{2}1$} &
\multicolumn{5}{c}{$IJ=\frac{3}{2}1$} &
\multicolumn{5}{c}{$IJ=\frac{5}{2}1$} \\
\hline
$ij$ & $\langle H_{ij}^c \rangle_D$ & $\langle H_{ij}^c \rangle_E$ & $\langle H_{ij}^{cs} \rangle_D$ & $\langle H_{ij}^{cs} \rangle_E$
& ~~~ & $\langle H_{ij}^c \rangle_D$ & $\langle H_{ij}^c \rangle_E$ & $\langle H_{ij}^{cs} \rangle_D$ & $\langle H_{ij}^{cs} \rangle_E$
& ~~~ & $\langle H_{ij}^c \rangle_D$ & $\langle H_{ij}^c \rangle_E$ & $\langle H_{ij}^{cs} \rangle_D$ & $\langle H_{ij}^{cs} \rangle_E$ \\
\hline
12 & $-8/3$ & $-8/27$ & $-8/3$ & $-8/27$
&  & $-8/3$ & $-8/81$ & $-8/3$ & $-8/81$
&  & $-8/3$ & $8/27$ & $-8/3$ & $8/27$ \\

	13 & $-8/3$ & $-8/27$ & $-8/3$ & $-8/27$
	&  & $-8/3$ & $-8/81$ & $-8/3$ & $-8/81$
	&  & $-8/3$ & $8/27$ & $-8/3$ & $8/27$ \\
	
	14 & $0$ & $4/27$ & $0$ & $4/27$
	&  & $0$ & $4/81$ & $0$ & $4/81$
	&  & $0$ & $-4/27$ & $0$ & $-4/27$ \\
	
	15 & $0$ & $4/27$ & $0$ & $4/27$
	&  & $0$ & $4/81$ & $0$ & $4/81$
	&  & $0$ & $-4/27$ & $0$ & $-4/27$ \\
	
	16 & $0$ & $-8/27$ & $0$ & $-8/27$
	&  & $0$ & $-8/81$ & $0$ & $-8/81$
	&  & $0$ & $8/27$ & $0$ & $8/27$ \\
	
	23 & $-8/3$ & $-8/27$ & $-8/3$ & $-8/27$
	&  & $-8/3$ & $-8/81$ & $-8/3$ & $-8/81$
	&  & $-8/3$ & $8/27$ & $-8/3$ & $8/27$ \\
	
	24 & $0$ & $4/27$ & $0$ & $4/27$
	&  & $0$ & $4/81$ & $0$ & $4/81$
	&  & $0$ & $-4/27$ & $0$ & $-4/27$ \\
	
	25 & $0$ & $4/27$ & $0$ & $4/27$
	&  & $0$ & $4/81$ & $0$ & $4/81$
	&  & $0$ & $-4/27$ & $0$ & $-4/27$ \\
	
	26 & $0$ & $-8/27$ & $0$ & $-8/27$
	&  & $0$ & $-8/81$ & $0$ & $-8/81$
	&  & $0$ & $8/27$ & $0$ & $8/27$ \\
	
	34 & $0$ & $-8/27$ & $0$ & $-8/27$
	&  & $0$ & $-8/81$ & $0$ & $-8/81$
	&  & $0$ & $8/27$ & $0$ & $8/27$ \\
	
	35 & $0$ & $-8/27$ & $0$ & $-8/27$
	&  & $0$ & $-8/81$ & $0$ & $-8/81$
	&  & $0$ & $8/27$ & $0$ & $8/27$ \\
	
	36 & $0$ & $16/27$ & $0$ & $-304/27$
	&  & $0$ & $16/81$ & $0$ & $-304/81$
	&  & $0$ & $-16/27$ & $0$ & $304/27$ \\
	
	45 & $-8/3$ & $-8/27$ & $-8/3$ & $-8/27$
	&  & $-8/3$ & $-8/81$ & $-8/3$ & $-8/81$
	&  & $-8/3$ & $8/27$ & $-8/3$ & $8/27$ \\
	
	46 & $-8/3$ & $-8/27$ & $-8/3$ & $-8/27$
	&  & $-8/3$ & $-8/81$ & $-8/3$ & $-8/81$
	&  & $-8/3$ & $8/27$ & $-8/3$ & $8/27$ \\
	
	56 & $-8/3$ & $-8/27$ & $-8/3$ & $-8/27$
	&  & $-8/3$ & $-8/81$ & $-8/3$ & $-8/81$
	&  & $-8/3$ & $8/27$ & $-8/3$ & $8/27$ \\
	\hline
	\hline
\end{tabular}

\end{table*}
   where $\mathcal{A}'=9P_{36}$.~The indices $i$ and $j$ denote the $i$-th and $j$-th quarks, respectively. The matrix elements $\langle H_{ij}^{c}\rangle$, which contain the operator $\boldsymbol{\lambda_i^c}\cdot\boldsymbol{\lambda_j^c}$, reflect the factor relevant to the color-dependent interactions, including the confinement term and the coulomb part of the one-gluon exchange interaction. Meanwhile, the matrix elements $\langle H_{ij}^{cs}\rangle$, which include the spin-color operator $\boldsymbol{\lambda_i^c}\cdot\boldsymbol{\lambda_j^c} \boldsymbol{\sigma_i}\cdot\boldsymbol{\sigma_j}$, mainly characterize the spin-color factor associated with the chromomagnetic part of the one-gluon exchange interaction.  
    To save space, we present only the calculated matrix elements for the representative $\Delta\Sigma_b^*$ channel in Tables~\ref{matrix1}-\ref{matrix2}. The corresponding results for the remaining channels are not displayed, since they show the same qualitative behavior.

   As shown in Tables~\ref{matrix1}-\ref{matrix2}, the direct matrix elements $\langle H_{ij}^{c} \rangle_D$ and $\langle H_{ij}^{cs} \rangle_D$ for the quark pairs between the two clusters are zero, indicating that the direct color interaction between two color-singlet baryon clusters does not contribute to the baryon-baryon interaction. Therefore, the  behavior of these interactions is mainly generated through the exchange terms induced by the antisymmetrization operator. For different $I$ and $J$, the exchange matrix elements show sizable variations in both magnitude and sign, which leads to different strengths of the corresponding interaction terms.
   More explicitly, for the $J=0$ case shown in Table~\ref{matrix1}, all the 
$\langle H_{ij}^{cs} \rangle_E$ matrix elements in the $I=\frac{1}{2}$ state are negative, ranging from $-8/9$ to $-112/9$. This is consistent with the 
strong attractive one-gluon exchange contribution shown in Fig.~\ref{Figure8}(a). When the isospin increases to $I=\frac{3}{2}$, the corresponding $\langle H_{ij}^{cs} \rangle_E$ matrix elements are still negative, but their magnitudes are considerably reduced, with values ranging from $-8/27$ to $-112/27$. Therefore, the attractive contribution from the one-gluon exchange interaction becomes much weaker. For the highest-isospin $I=\frac{5}{2}$ state, the $\langle H_{ij}^{cs} \rangle_E$ matrix elements become positive, ranging from $8/9$ to $112/9$. Correspondingly, the one-gluon exchange interaction becomes repulsive, as shown in Fig.~\ref{Figure8}(c).

    Therefore, the matrix elements clarify the microscopic origin of the different one-gluon exchange contributions in different isospin channels. The weakening and eventual sign reversal of $\langle H_{ij}^{cs} \rangle_E$ with increasing isospin explain why the one-gluon exchange potential changes from strongly attractive in the $I=\frac{1}{2}$ state, to weakly attractive in the $I=\frac{3}{2}$ state, and finally to repulsive in the $I=\frac{5}{2}$ state.

  \label{Resonance states}

	\section{Summary}
	\label{22}
	In this work, we have systematically investigated all possible $S$-wave singly bottomed dibaryon states in the chiral quark model, with isospin $I=0,~\frac{1}{2},~1,~\frac{3}{2},~2,~\frac{5}{2}$ and the total angular momentum $J=0,~1,~2,~3$. Our work mainly focuses on the calculation of binding energies and the study of interaction between two baryons.
    %The resonating group method and the generating coordinate method are employed to solve the six-quark dynamics, and all possible $S$-wave baryon-baryon channels with different isospin and total angular momentum are considered. 
    
    The bound state calculations show that several channels can support bound solutions, including $\Delta\Sigma_b^*$, $\Delta\Sigma_b$, $\Sigma\Sigma_b$, $\Sigma\Sigma_b^*$, $\Sigma^*\Sigma_b$, and $\Sigma^*\Sigma_b^*$ systems.
    Among these, the $\Delta\Sigma_b^*$ system is found to be bound state for $IJ=\frac{1}{2}0,~\frac{1}{2}1~,\frac{1}{2}2,~\frac{1}{2}3,\frac{3}{2}3$ with binding energies of -32.93, -38.81, -36.96, -35.93, and -12.25 MeV, respectively. The $\Delta\Sigma_b$ system exhibits bound states for the quantum number $IJ=\frac{1}{2}1,~\frac{3}{2}1,~\frac{1}{2}2,~\frac{3}{2}2$, with corresponding binding energies of -38.43, -0.31, -28.68, and -1.55 MeV, respectively. The $\Sigma^*\Sigma_b^*$ system supports bound states for $IJ=00,~01,~02,~03,~13$, and the binding energies are -33.68, -24.16, -19.62, -24.21, and -2.27 MeV, respectively. The binding energies for the remaining systems are given in the Tables \ref{bound energy1}-\ref{bound energy2}.
     The calculation of binding energies also reveals a useful but not absolute pattern related to the baryon multiplet structure. Channels composed of two decuplet baryons, such as $\Delta\Sigma_b^*$ and $\Sigma^*\Sigma_b^*$, are generally more favorable for binding. Channels formed by one decuplet baryon and one octet baryon, such as $\Delta\Sigma_b$ and $\Sigma^*\Sigma_b$, can also support bound states, although the attraction is usually weaker than that in decuplet-decuplet systems. In contrast, channels composed of two octet baryons are generally less favorable for binding. Nevertheless, this tendency is not universal, as some octet-octet channels may also generate bound solutions. Therefore, a systematic calculation over all possible quantum number channels is essential for identifying the most promising singly bottomed dibaryon candidates.
     
     To further understand the interaction mechanism behind these binding patterns, we analyze the effective potentials of the representative channels. The behavior of the effective potential is compatible with the bound state calculations. Several features can be identified. For a fixed total angular momentum $J$, the low-isospin channels usually exhibit deeper attractive potentials, while increasing the isospin weakens the attraction and eventually makes the system unbound. However, for a fixed isospin, the dependence on $J$ is not universal. In some systems, the potential of lower-$J$ shows more attractive behavior, whereas in others the higher-$J$ state can be more favorable. A similar situation also appears in hidden-charm systems $P_c$. Therefore, the binding behavior of singly bottomed dibaryons cannot be determined by either $I$ or $J$ alone, but is governed by the combined effects of isospin-spin coupling and the internal baryon structure.
    
   To clarify the microscopic origin of the attraction, we further analyze the decomposed effective potentials and the corresponding matrix elements in the spin-flavor-color space. The results show that the scalar $\sigma$ meson exchange provides the dominant intermediate-range attraction, while the one-gluon exchange interaction plays an important role in the short-range region. 
Depending on the quantum numbers, the one-gluon exchange contribution can be attractive or repulsive, which is closely related to the exchange matrix elements induced by the antisymmetrization operator. 
     %The kinetic energy term also gives non-negligible contributions and may become effectively attractive in some channels. 
     Thus, the formation of total effective potential is the result of cancellation among the $\sigma$ meson exchange, the one-gluon exchange interaction, the kinetic-energy contribution, and the quark-exchange effect.
    
   The present study provides useful guidance for future experimental searches. 
   Our results indicate that the low-isospin channels are more favorable for binding. Meanwhile, a tendency can be observed that decuplet-decuplet systems support relatively more bound states.
    Therefore, the low-isospin configurations and decuplet-decuplet systems may serve as promising targets in future searches for singly bottomed dibaryons. To provide more detailed information for experimental investigations, it is also necessary to further explore their possible decay properties. 
   For example, the $\Delta\Sigma_b^*$ dibaryon may decay into the $N\Sigma_b^*$, $N\Sigma_b$, and $N\Lambda_b$ channels. Similarly, the $\Sigma^*\Sigma_b^*$ dibaryon may decay into the $N\Xi_{b}^{\prime}$, $N\Xi_{b}$, and $\Lambda \Lambda_{b}$ channels.
In future work, we will investigate the scattering phase shifts of these decay channels to further determine whether these states are resonances.

\acknowledgments{This work is supported partly by the National Natural Science Foundation of China under Contracts Nos. 12575088, 11675080, 11775118 and 11535005.}

\end{document}